\documentclass[12pt]{article}
\usepackage{epsfig,amssymb,amsmath,psfrag,subfigure}

\def \be  {\begin{equation}}
\def \ee  {\end{equation}}
\def \ba  {\begin{eqnarray}}
\def \ea  {\end{eqnarray}}
\def \baa {\begin{eqnarray*}}
\def \eaa {\end{eqnarray*}}
\def \lab #1 {\label{#1}}

\newcommand\re[1]{(\ref{#1})}

\def \matrix #1 {\left(\begin{array}{cc} #1 \end{array}\right)}

\def \tr {\mathop{\rm tr}\nolimits}

\def \e  {\mathop{\rm e}\nolimits}
\newcommand\lr[1]{{\left({#1}\right)}}

\newcommand \vev [1] {\langle{#1}\rangle}
\newcommand \VEV [1] {\left\langle{#1}\right\rangle}
\newcommand \ket [1] {|{#1}\rangle}
\newcommand \bra [1] {\langle {#1}|}

\newcommand{\ft}[2]{{\textstyle\frac{#1}{#2}}}
\begin{document}

\begin{titlepage}

\thispagestyle{empty}

\null\vskip-43pt \hfill
\begin{minipage}[t]{45mm}
IPhT--T11/034 \\ CERN-PH-TH/2011-055
\end{minipage}

\vspace*{3cm}

\centerline{\large \bf Are scattering amplitudes dual to super Wilson loops?}
\vspace*{1cm}

\centerline{\sc A.V. Belitsky$^a$,  G.P. Korchemsky$^b$, E. Sokatchev$^{c,d,e}$}

\vspace{10mm}

\centerline{\it $^a$Department of Physics, Arizona State University}
\centerline{\it Tempe, AZ 85287-1504, USA}

\vspace{3mm}

\centerline{\it $^b$Institut de Physique Th\'eorique\footnote{Unit\'e de Recherche Associ\'ee au CNRS URA 2306}, CEA Saclay}
\centerline{\it 91191 Gif-sur-Yvette Cedex, France}

\vspace{3mm}
\centerline{\it $^c$ Physics Department, Theory Unit, CERN}
\centerline{\it CH -1211, Geneva 23, Switzerland}

\vspace{3mm}
\centerline{\it $^d$ Institut Universitaire de France}
\centerline{\it 103, bd Saint-Michel
F-75005 Paris, France}

\vspace{3mm}
\centerline{\it $^e$ LAPTH\,\footnote[2]{Laboratoire d'Annecy-le-Vieux de Physique Th\'{e}orique, UMR 5108},   Universit\'{e} de Savoie, CNRS}
\centerline{\it  B.P. 110,  F-74941 Annecy-le-Vieux, France}

\vspace{1cm}

\centerline{\bf Abstract}

\vspace{5mm}

The MHV scattering amplitudes in planar $\mathcal{N}=4$ SYM are dual to 
bosonic light-like Wilson loops. We explore various proposals for extending this 
duality to generic non-MHV amplitudes. The corresponding dual object should 
have the same symmetries as the scattering amplitudes and be invariant 
to all loops under the chiral half of the $\mathcal{N}=4$ superconformal 
symmetry. We analyze the recently introduced supersymmetric extensions of the 
light-like Wilson loop (formulated in Minkowski space-time) and demonstrate that 
they have the required symmetry properties at the classical level only, up to terms 
proportional to field equations of motion. 
At the quantum level, due to the specific light-cone singularities of the Wilson loop, the
equations of motion produce a nontrivial finite contribution which breaks some of the
classical symmetries. As a result, the quantum corrections violate the chiral 
supersymmetry already at one loop, thus invalidating the conjectured duality between
Wilson loops and non-MHV scattering amplitudes. We compute
the corresponding anomaly to one loop and solve the supersymmetric
Ward identity to find the complete expression for the rectangular Wilson loop at 
leading order in the coupling constant. We also demonstrate that this result is
consistent with conformal Ward identities by independently evaluating corresponding
one-loop conformal anomaly.
 
\end{titlepage}

\setcounter{footnote} 0

\newpage

\pagestyle{plain}
\setcounter{page} 1

\section{Introduction}

Recent studies revealed that scattering amplitudes in planar $\mathcal{N}=4$ supersymmetric 
Yang-Mills theory (SYM) possess a new hidden symmetry \cite{DrunHenKorSok10,BerMal08}. The 
latter appears in addition to the conventional superconformal symmetry of the Lagrangian of the 
theory and leads to powerful constraints on the form of the all-loop scattering amplitudes. 

A distinguishing feature of $\mathcal{N}=4$ SYM is that all on-shell states  (gluons with helicity $\pm 1$, 
gluinos with helicity $\pm 1/2$ and scalars) can be encoded in a single superstate $\Phi(p_i,\eta_i)$
\cite{BriMan82,Nair}. States with different helicities appear as coefficients in the expansion of the superstate in powers of the odd 
variables $\eta_i^A$ (with $A=1,2,3,4$).  As a consequence,  all $n-$particle scattering amplitudes can 
be combined into a single superamplitude $\mathcal{A}_n$. Supersymmetry restricts the form of the 
superamplitude to be
\begin{align}\label{dec}
\mathcal{A}_n = \mathcal{A}_n^{\rm MHV} + \mathcal{A}_n^{\rm NMHV} 
+ \ldots + \mathcal{A}_n^{\rm N^{n-4}MHV}\,,
\end{align}
where the $k$th term in the sum, $\mathcal{A}_n^{\rm N^kMHV}$, describes the scattering amplitudes with 
total helicity of the particles $(-n+4+2k)$. It has the following general form  \cite{Nair}
\begin{align}\label{gen}
\mathcal{A}_n^{\rm N^kMHV} 
= i (2\pi)^4
\frac{\delta^{(4)}(p^{\dot\alpha\alpha})\delta^{(8)}(q^A_\alpha)}{\vev{12}\vev{23}\ldots \vev{n1}}
\, 
\widehat{\mathcal{A}}_{n;k}(\lambda,\tilde\lambda,\eta;a)\,,
\end{align}
where $p^{\dot\alpha\alpha}=\sum_i p_i^{\dot\alpha\alpha}$ and $q^{\alpha\,A}=\sum_i 
\lambda_i^\alpha\eta_i^A$ 
is the total momentum and chiral supercharge of the $n$ particles. Here we used the 
spinor-helicity formalism to parameterize the light-like momenta  in terms of two-component commuting spinors 
$p_j^{\dot\alpha\alpha} = \widetilde\lambda^{\dot\alpha}_j \lambda_j^\alpha$ and the angular brackets $\vev{jk}$ 
are defined in Appendix A. The dependence of the scattering amplitude  on the 't Hooft coupling constant 
$a=g^2 N_c/(4\pi^2)$ is carried by the nontrivial function $\widehat A_{n;k}(\lambda,\tilde\lambda,\eta;a)$
which is given by a homogenous polynomial in the $\eta$'s of 
degree $(4k)$. At tree level, the function $\widehat A_{n;k}(\lambda,\tilde\lambda,\eta)$ enjoys the full 
$PSU(2,2|4)$ superconformal symmetry of the $\mathcal{N}=4$ SYM, while at loop level
the dilatations, the special conformal boosts and their supersymmetric extension are broken by infrared divergences \cite{Witten:2003nn}.

As we already mentioned, the superamplitude in planar $\mathcal{N}=4$ SYM has another, hidden 
symmetry. This symmetry, called ``dual superconformal symmetry" in Ref.~\cite{DrunHenKorSok10}, acts naturally on the so-called 
region supermomenta  $(x_i,\theta_i)$ defined as 
\begin{align}\label{super-mom}
p_i^{\dot\alpha\alpha}=\widetilde\lambda^{\dot\alpha}_i \lambda_i^\alpha
=
(x_i-x_{i+1})^{\dot\alpha\alpha}
\,,\qquad 
\lambda_i^\alpha \,\eta_i^A = (\theta_i - \theta_{i+1})^{\alpha\, A}\,.
\end{align}
In spite of the fact that $(x_i,\theta_i)$  have the meaning of momenta, the dual symmetry acts on 
them as if they were coordinates in configuration (rather than momentum) space 
\cite{B93,DruHenSmiSok06,DrunHenKorSok10}. At tree level, the superamplitude 
$\mathcal{A}_n^{(0)}$ enjoys  exact dual superconformal symmetry, while at loop level some of the dual 
symmetries become anomalous (see below). 
 
Dual superconformal symmetry emerges as a hidden property of the scattering amplitudes in planar 
$\mathcal{N}=4$ SYM. For the simplest, maximally helicity violating superamplitude 
$\mathcal{A}_n^{\rm MHV}$, the dual conformal symmetry becomes manifest through the conjectured 
duality between the functions $\widehat{\mathcal{A}}_{n;0}$ defining the perturbative loop corrections to the MHV 
superamplitude, Eq.~\re{gen}, and light-like Wilson loops  $W_n$
\footnote{The duality between planar amplitudes and light-like Wilson loops also holds in gauge theories 
with less or no supersymmetry, including QCD. However, in distinction with $\mathcal{N}=4$ SYM, there the
relation \re{MHV-dual} holds in the high-energy (Regge) limit only~\cite{Korchemskaya:1996je,KorDruSok08}.}
\begin{align}\label{MHV-dual}
\ln  \widehat{\mathcal{A}}_{n;0} = \ln W_n +O(\varepsilon)  \,,\qquad
W_n= \frac{1}{N_c} \VEV{ \tr P\exp\lr{i g\int_{C_n} dx\cdot A(x)}}
\, ,
\end{align}
where $O(\varepsilon)$ stands for terms vanishing in dimensional regularization and the Wilson loop is 
evaluated along the polygon light-like contour $C_n=[x_1, x_2] \cup [x_2, x_3] \dots [x_n, x_1]$ formed 
by the particle momenta $p_i=x_i-x_{i+1}$.
The power of the duality relation \re{MHV-dual} exhibits itself 
through the fact that the dual conformal symmetry of the MHV superamplitude is mapped into the conventional 
conformal symmetry of the Wilson loop in $\mathcal{N}=4$ SYM. First hinted at
in Ref.~\cite{AldMal08} by the 
strong coupling analysis for $n=4,5$ via the AdS/CFT correspondence, 
the duality between the two objects was understood as the invariance of the string sigma model on  an  
AdS$_5 \times$S$^5$ background under $T$-duality transformations for the bosonic \cite{Kallosh:1998ji} and fermionic variables 
\cite{BerMal08}. At weak coupling the proposal (\ref{MHV-dual}) was put forward in Ref.\ \cite{KorDruSok08} and 
confirmed at one loop for an arbitrary number of cusps $n$ in Ref.~\cite{BraHesTra07} and at two loops 
for $n=4,5-$point Wilson loops \cite{KorDruHenSok08,DruHenKorSok07}  by confronting these predictions with the 
available results for gluon scattering amplitudes \cite{ABDK03,BDS05}. At the moment, the most thorough test of the 
duality \re{MHV-dual} comes from the comparison of  the two-loop results for the $n=6$ (hexagon) Wilson loop 
\cite{DruHenKorSok08,AnaBraHesKhoSpeTra09,DelDuhSmi10,GonSprVerVol10} with the six-gluon MHV scattering 
amplitude computed to the same order in the coupling \cite{BerDixKosRoiSprVerVol08,Cachazo:2008hp}.

The MHV superamplitude is the simplest among all superamplitudes in \re{dec}. In particular, the 
corresponding function $\widehat{\mathcal{A}}_{n;0}$ defined in Eq.\ \re{gen} only depends on the bosonic 
variables and is independent of the odd $\eta-$variables. At tree level, it is given by 
$\widehat{\mathcal{A}}_{n;0}=1+O(a)$ and, according to the duality relation \re{MHV-dual}, its loop corrections 
coincide (upon appropriate identification of the regularization parameters) with those of the bosonic light-like 
Wilson loop $W_n$. A natural question arises whether 
the duality relation \re{MHV-dual} can be extended to the full superamplitude $\mathcal{A}_n$.  If 
such a duality relation exists, the bosonic Wilson loop in the right-hand side of \re{MHV-dual} should 
appear as the first term in the expansion of the dual object $\mathcal{W}_n$ in powers of the
$\eta-$variables analogous to \re{dec}. Viewed as a function of the dual coordinates $(x_i,\theta_i)$, it 
should possess the same symmetry as the superamplitude. Namely, the duality between $\mathcal{A}_n$ 
and $\mathcal{W}_n$ would imply that the dual superconformal symmetry of the former follows from 
the conventional superconformal symmetry of the latter in $\mathcal{N}=4$ SYM. Moreover, since 
some of the symmetries of the superamplitudes are broken at the quantum level,  $\mathcal{W}_n$ should 
have the same anomalies. 
Regarding the $\mathcal{N}=4$ supersymmetry, the dual object  
$\mathcal{W}_n$ should have the following unusual property -- it depends only on the chiral $\theta-$variables 
but not on $\bar\theta$. The reason for this can be traced back to the chiral formulation of scattering 
amplitudes in the on-shell superspace formalism. In addition, a serious complication is that $\mathcal{N}=4$ SYM 
only exists on shell, i.e. the algebra of the supersymmetry transformations closes modulo field equations. This makes 
the construction of  $\mathcal{W}_n$ somewhat tricky and, as we will see later in the paper, leads to various subtleties.  

Recently, two different proposals for $\mathcal{W}_n$ were put forward. In Ref.\ \cite{MasSki10}  Mason and Skinner put forward a 
formulation of  $\mathcal{W}_n$ in twistor space.
Yet another form of $\mathcal{W}_n$, this time  in Minkowski space, was suggested by Caron-Huot in Ref.\ \cite{Car10}.  It was claimed in 
\cite{MasSki10} that all amplitudes in $\mathcal{N}=4$ SYM are described by a supersymmetric Wilson loop in twistor space and the 
same statement was apparently implied for its Minkowski transform. The Minkowski version takes the following form \cite{OogRahRobTan00}
\begin{align}\label{super-W}
 \mathcal{W}_n =  \frac{1}{N_c}\VEV{ \tr P\exp\lr{ig \int_{\mathcal{C}_n} dx^{\mu} \mathcal{A}_{\mu} (x,\theta)
+
ig \int_{\mathcal{C}_n} d \theta^{\alpha A} \mathcal{F}_{\alpha A}(x,\theta)}}\,,
\end{align}
where the integration goes over a contour ${\mathcal{C}_n}$ in superspace formed by $n$ straight segments 
connecting the points $(x_i,\theta_i)$ and defined by the supermomenta of the scattered particles \re{super-mom}. 
Here $\mathcal{A}_{\mu}$ and $\mathcal{F}_{\alpha A}$ are the bosonic and fermionic superspace gauge connections,
respectively, 
given by a $\theta-$expansion whose coefficients are the various component fields (gluon, gauginos and scalars) of $\mathcal{N}=4$ SYM. 
Suppressng the dependence on the $\theta-$variables, the supersymmetrized Wilson loop \re{super-W} 
reduces to the bosonic Wilson loop $W_n$ in \re{MHV-dual}. Another proposal for $\mathcal{W}_n$ 
was presented in \cite{Car10} where certain perturbative data and BCFW-like recursion relations  \cite{ArkaniHamed:2010kv,Boels:2010nw}
were used 
to advocate a particular supersymmetric generalization of the bosonic light-like Wilson loop. The resulting
supersymmetric Wilson loop was argued to be equivalent to \re{super-W}. Thus, both proposals suggest 
the duality ${\mathcal{A}}_n \sim \mathcal{W}_n$.

By construction, the supersymmetric Wilson loop $\mathcal{W}_n$ automatically reproduces the MHV 
amplitudes through the 
duality relation \re{MHV-dual}. In addition, it yields a definite prediction for non-MHV amplitudes to all loops.
In this paper, we verify the duality between the superamplitudes $\mathcal{A}_n$ and the supersymmetric 
Wilson loop $\mathcal{W}_n$ by performing one-loop calculations of \re{super-W} in $\mathcal{N}=4$ SYM. 
We show by 
an explicit one-loop computation that, contrary to the above expectations, the perturbative corrections to $\mathcal{W}_n$ do 
not match those of the non-MHV amplitudes, thus invalidating the duality relation $\mathcal{A}_n\sim\mathcal{W}_n$
beyond the MHV amplitudes. We show that the reason for this has to do with the fact that $\mathcal{W}_n$ is 
invariant only under on-shell chiral $\mathcal{N}=4$ supersymmetric transformations, that is up to contributions 
proportional to the equations of motion.
At loop level, due to the specific light-cone singularities of the Wilson loops \cite{KK92}, the equations
of motion generate a nontrivial finite effect. As a result, the supersymmetry of $\mathcal{W}_n$ gets 
broken by quantum corrections. We compute the corresponding anomaly to one-loop order and demonstrate 
that the result obtained for $\mathcal{W}_n$ is in a perfect agreement with the anomalous supersymmetric Ward 
identities.  

An alternative proposal for the dual description of scattering amplitudes in planar
$\mathcal{N}=4$ SYM via the light-cone limit of correlation functions
of certain half-BPS operators was put forward in Refs.\ \cite{EdeKorSok10,EHKS}.   It proves to be immune to the 
problems that one encounters in the construction of the super Wilson loops, as we discuss
in the Conclusions.

The paper is organized as follows. After a discussion of the basic symmetries that any
proposal for the dual description of superamplitudes should fulfill,  in Sect.~2 we introduce the supersymmetric 
generalizations of the bosonic Wilson loop proposed in Refs.~\cite{MasSki10,Car10}  and explicitly 
demonstrate that the two differ by the
presence of equations of motion when considered off shell. Then, in Sect.\ \ref{SUSYWI}, we derive the 
Ward identities resulting from the the transformation properties of  the
Wilson loops under   Poincar\'e supersymmetry and calculate the corresponding anomalies. In Sect.\ 
\ref{OneLoopWL} we solve the anomalous Ward identities and reconstruct the explicit form of the one-loop 
correction to the rectangular Wilson loop $\mathcal{W}_4$. Sect.\ \ref{Conclusions} contains
concluding remarks. Details on our conventions and normalizations are given in the Appendix A. Finally, the 
Ward identities corresponding to the conformal transformations of supersymmetric Wilson loop are discussed 
in Appendix B, where we demonstrate their complete consistency with our findings in the main text.

\section{Supersymmetric Wilson loop}

As  already reviewed in the Introduction, the symmetries of the scattering amplitudes 
extend far beyond the conventional  superconformal symmetry of $\mathcal{N}=4$ SYM. 
If a generalization of the Wilson loop dual to superamplitudes exists, 
both objects will possess the same symmetries. This implies that the dual superconformal 
symmetry of the amplitudes should match the conventional superconformal $
\mathcal{N}=4$  symmetry of the supersymmetric Wilson loop.

\subsection{Matching the symmetries of the $\mathcal{N}=4$  amplitudes}
\label{Matching}

Let us assume for a moment that the $\mathcal{N}=4$ superamplitudes are indeed dual to $\mathcal{W}_n$ 
and let us summarize the constraints imposed by the dual superconformal symmetry of the amplitude 
$\mathcal{A}_n$ on the properties of $\mathcal{W}_n$.

By construction,  $\mathcal{W}_n$ depends on the Grassmann variables $\theta_i^A$ carrying   $SU(4)$
charges. Then by virtue of $SU(4)$ invariance, it admits an expansion in powers $(\theta)^{4k}$:
\be\label{W-exp}
\mathcal{W}_n  = \mathcal{W}_{n;0}(x_i;a)+ \mathcal{W}_{n;1}   (x_i,\theta_i^A;a)
+ \ldots +\mathcal{W}_{n;n}(x_i,\theta_i^A;a) 
\, .
\ee
Here $\mathcal{W}_{n;k}$ is a homogeneous $SU(4)$ invariant polynomial of degree $4k$ in the Grassmann 
variables $\chi_i^A = \vev{ i \theta_i^A}$ (see Eq.~\re{EtaToChi} below) and the expansion runs up to $k=n$. The lowest term 
of the expansion does not depend on the odd variables and, therefore, it coincides with the bosonic Wilson 
loop  $\mathcal{W}_{n;0}(x_i;a) = W_n$, (see Eq.~\re{MHV-dual}).

The expansion \re{W-exp} is similar to that of the scattering amplitude \re{dec}. In fact, the conjectured duality 
between superamplitudes and supersymmetric Wilson loops establishes the correspondence between the two 
expansions~\cite{MasSki10,Car10}
\begin{align}\label{dual-par}
 \mathcal{W}_{n;k}   (x,\theta;a) =a^k \widehat{\mathcal{A}}_{n;k}(\lambda,\tilde\lambda,\eta;a)   \,,
\end{align}
where the functions $ \widehat{\mathcal{A}}_{n;k}$ define the perturbative corrections to the N${}^k$MHV 
superamplitude, 
Eq.~\re{gen}. For $k=0$ the relation \re{dual-par} reproduces the duality between the MHV amplitude and the
bosonic Wilson loop, Eq.~\re{MHV-dual}. Notice the appearance of a power of the coupling constant in the right-hand side of 
\re{dual-par}.%
\footnote{Following  \cite{Car10}, the factor $a^k$ in the right-hand side of \re{dual-par} can be eliminated through a redefinition
of the odd variables $\eta\to a^{-1/4}\eta$.}
The reason for it is that  the perturbative expansion of 
$\mathcal{W}_{n;k}$  starts at order $O(a^k)$. Then, the duality relation \re{dual-par} leads to the following iterative 
structure of loop corrections:  Having computed $\mathcal{W}_n$ to, say, $k-$loops and expanding the result in 
powers of $\theta$'s as in \re{W-exp}, we will be able to determine the $k-$loop corrections to the MHV amplitudes, the
$(k-1)-$loop corrections to the NMHV amplitudes and so on until we reach the tree-level expression for 
N${}^k$MHV amplitudes.

We observe that, for a given number of external particles $n$, the expansion in \re{gen}terminates at  $k=n-4$ and, 
therefore, $\widehat{\mathcal{A}}_{n;k}=0$ for $k\ge n-3$. This property is an immediate consequence of the 
conventional on-shell Poincar\'e supersymmetry of the all-loop amplitudes in $\mathcal{N}=4$ SYM. Then, the
duality relation \re{dual-par} implies that the top four  components of the expansion \re{W-exp} should vanish 
to all loops
\begin{align}\label{W=0}
\mathcal{W}_{n;n-3}(x,\theta;a) 
=  \mathcal{W}_{n;n-2}(x,\theta;a) 
= \mathcal{W}_{n;n-1}(x,\theta;a) 
= \mathcal{W}_{n;n} (x,\theta;a)\stackrel{?}{=}  0\,.
\end{align}
We recall that the perturbative expansion of $\mathcal{W}_{n;n-3}(x,\theta;a)$ only starts at $(n-3)-$loops and 
accordingly for the remaining functions in this relation. Here we inserted the question
mark (see also Eqs.~\re{W-sym}, \re{super-W4} and \re{KW=0} below) to indicate that these relations
have the status of a conjecture. Below we shall check them by explicit one-loop
calculations.
 
Tree-level amplitudes are free from infrared divergences and, as a result, they enjoy both conventional and 
dual superconformal symmetries. Then, the duality relation \re{dual-par} predicts that the lowest $O(a^k)$ 
correction to $\mathcal{W}_{n;k}   (x,\theta;a)$ should respect the dual superconformal symmetry. At loop 
level, the scattering amplitudes suffer from infrared divergences and some of the symmetries become anomalous. 
It is paramount for our purposes that the dual $Q-$ and $\bar S-$supersymmetries remain unbroken to all 
loops. The reason for this is as follows. As it is evident from its definition, the $Q-$supersymmetry generates
shifts of the odd variables $\delta_Q \theta_i{}_\alpha^{A}=\epsilon_\alpha^{A}$. Since the scattering amplitudes
depend on $\theta$'s only through the differences \re{super-mom}, they stay invariant under the action of 
$Q-$supersymmetry. In a similar manner, the dual $\bar S-$supersymmetry coincides with the conventional 
$\bar q-$supersymmetry of the $\mathcal{N}=4$ Lagrangian. As long as we employ a regularization that 
preserves Poincar\'e supersymmetry, $\bar q-$supersymmetry remains unbroken and the same should be true for the dual $\bar S-$supersymmetry, $\delta_{\bar S} (\theta_i)_\alpha^{A}=
(x_i)_{\alpha\dot\alpha}\xi^{\dot\alpha\, A}$. 

Thus, the duality relation \re{dual-par} implies 
that the supersymmetric Wilson loop should preserve $Q-$ and $\bar S-$supersymmetries to all loops
\begin{align}\label{W-sym}
 Q_A^\alpha \, \mathcal{W}_n(x,\theta;a) =\bar S_{\dot\alpha\, A} \,  \mathcal{W}_n(x,\theta;a) \stackrel{?}{=} 0\,.
\end{align}
To discuss the consequence of these relations, it is convenient to introduce new variables \cite{hodges}, the 
so-called momentum supertwistors $Z^a=(\lambda_{i}, \mu_i \,, \chi_i)$ with 
\begin{align}\label{twistors}
\chi_i^A = \vev{i\theta_i^A}\,,\qquad \mu_{i \,\dot\alpha} = \lambda_i^\alpha(x_i)_{\alpha\dot\alpha}  \,.
\end{align}
The dual superconformal symmetry acts on $Z^a$ as linear $GL(4|4)$ transformations. In 
particular, the action of the dual $Q-$ and $\bar S-$supersymmetries corresponds to
\begin{align}\label{chi-rot}
  \chi_i^A \to \chi_i^A + \vev{ i \epsilon^A} + [ \mu_i\,\bar \xi^{A}]\,,
\end{align}
with $\epsilon_\alpha^{A}$ and $\bar \xi^{\dot\alpha A}$ being the parameters of the corresponding
transformations. Using the invariance of the supersymmetric Wilson loop \re{W-sym}
under the transformations \re{chi-rot}, we can always choose the parameters $\epsilon^A_\alpha$ 
and $\xi^{\dot\alpha \, A}$ in such a manner 
as to set 16 components of the Grassmann variables to zero, e.g., 
$
\chi_1^A=\chi_2^A=\chi_3^A=\chi_4^A=0
$.
This immediately implies that the supersymmetric Wilson loop depends on $(n-4)$ odd variables $\chi_i$ 
and, therefore, its expansion \re{W-exp}  terminates at the term $ \mathcal{W}_{n;n-4}$, in agreement with 
\re{W=0}. In particular, in the special case of $n=4$, the supersymmetric Wilson 
loop $\mathcal{W}_4$ cannot depend on the odd variables and, therefore, it should coincide with the 
bosonic Wilson loop
\begin{align}\label{super-W4}
 \mathcal{W}_4(x,\theta;a) \stackrel{?}{=}  W_4(x;a)\,.
\end{align}
This is in accord with the well-known fact that for $n=4$ particles the only nonzero amplitude in 
$\mathcal{N}=4$ SYM is the MHV one. 
 
Infrared divergences break the dual conformal symmetry of the amplitudes making special
conformal $K-$symmetry anomalous, $K^{\dot\alpha\alpha}\widehat{\mathcal{A}}_{n;k}\neq 0$. Since the 
infrared divergences have a universal form independent of the helicity configuration of the scattered particles, they 
cancel in the ratio of amplitudes $\widehat{\mathcal{A}}_{n}/\widehat{\mathcal{A}}_{n;0}$. As a consequence, dual conformal symmetry gets restored in the ratio, 
$K^{\dot\alpha\alpha}\big({\widehat{\mathcal{A}}_{n}/\widehat{\mathcal{A}}_{n;0}}\big)= 0$.%
\footnote{This property has been conjectured in Ref.~\cite{DrunHenKorSok10} and 
was recently verified in Ref.~\cite{KosRoiVer10} by an explicit two-loop calculation of the 
$n=6$ NMHV amplitude in $\mathcal{N}=4$ SYM.} Together with the duality
relation \re{dual-par} this leads to the following constraint on the supersymmetric Wilson loop
\begin{align}
\label{KW=0}
 K^{\dot\alpha\alpha}\lr{ {\mathcal{W}}_{n}/ \mathcal{W} _{n;0}}\stackrel{?}{=}  0\,.
\end{align}
In other words, the ratio of the supersymmetric Wilson loop and its lowest bosonic component should be invariant under the special conformal transformations in $\mathcal{N}=4$ SYM.

\subsection{Supersymmetric connections}
\label{SUSYconnection}

Let us start with reviewing the construction of the supersymmetric Wilson loop \re{super-W} 
introduced in Refs.\ \cite{MasSki10,Car10}. The super Wilson loop in  \re{super-W} is given 
by the path-ordered product of supersymmetric Wilson lines calculated along the 
segments of the polygon contour $\mathcal{C}_n$ 
\begin{align}\label{W-prod}
\mathcal{W}_n = \frac1{N_c} \vev{ \tr  \lr{ \mathcal{W}_{[1,2]}  \mathcal{W}_{[2,3]}\ldots  \mathcal{W}_{[n,1]}} }  \,.
\end{align}
Here the Wilson line $\mathcal{W}_{[i,i+1]}$ is evaluated along the straight line
connecting the points $(x_i,\theta_i)$ and $(x_{i+1},\theta_{i+1})$ in the chiral superspace
\begin{align}\label{super-path}
x(t_i) = x_i- t_i x_{i,i+1}\,,\qquad \theta(t_i) = \theta_i -t_i \theta_{i,i+1}\,,\qquad
(0\le t_i \le 1)\,,
\end{align}
with implied periodicity conditions $x_{i+n}\equiv x_i$ and $\theta_{i+n}\equiv \theta_i$.
Then, the supersymmetric Wilson line takes the form
\be\label{W-link}
\mathcal{W}_{[i,i+1]} = T\exp \left( ig \int_0^1 dt_i\, \mathcal{B}(t_i) \right)\,,
\ee
where the (M)inkowski (S)uperspace connection $\mathcal{B}(t_i)$ is the sum of the bosonic and fermionic  connections
projected on the tangent direction to the trajectory,
\be
\label{Bsuperfield}
\mathcal{B}^{\rm MS}(t_i)
= 
\frac12 \dot x^{\dot\alpha\alpha}(t_i) {\mathcal A}_{\alpha\dot\alpha}
+
\dot \theta^{\alpha A}(t_i){\mathcal F}_{\alpha A} 
\, .
\ee
The supersymmetric Wilson loop \re{W-prod} is invariant under non-Abelian gauge transformations
\begin{align}\label{B-gauge}
& \mathcal{B}(t_i) \to U^\dagger(t_i) \mathcal{B}(t_i) U(t_i) + \frac{i}{g} U^\dagger \partial_{t_i} U\,, 
\qquad
  \mathcal{W}_{[i,i+1]} \to U^\dagger(1) 
\mathcal{W}_{[i,i+1]} U(0)\,,
\end{align}
with $U=U(t_i)$.
For the bosonic and fermionic connections the gauge transformations 
look as
\begin{align}\label{gauge}
{\mathcal A}_{\alpha\dot\alpha} &\to U^\dagger {\mathcal A}_{\alpha\dot\alpha} U
+ \frac{i}{g} U^\dagger\partial_{\alpha\dot\alpha} U\,,\qquad
\\[2mm] \notag
{\mathcal F}_{\alpha A} &\to U^\dagger {\mathcal F}_{\alpha A}U  +  \frac{i}{g} U^\dagger  \partial_{\alpha A} U
\,,
\end{align}
where $\partial_{\alpha\dot\alpha}\equiv \partial/\partial {x^{\dot\alpha\alpha}}$ and 
$\partial_{\alpha A} \equiv \partial/\partial \theta^{\alpha A}$.

Let us now present the explicit expressions for the bosonic, $\mathcal{A}(x,\theta)$,
and fermionic, $\mathcal{F}(x,\theta)$, connections.
Both of them are given by expansions in the Grassmann variable $\theta^{A}_\alpha$ with 
the coefficients of increasing scaling dimension built from the various field components 
(gluon $A^{\dot\alpha\alpha}$, gauginos $\psi^A_\alpha, \bar\psi_A^{\dot\alpha}$ and 
scalars $\phi^{AB}, \bar\phi_{AB} = \ft12 \epsilon_{ABCD} \phi^{CD}$) and their gauge covariant 
derivatives $D_{\alpha\dot\alpha}=\partial_{\alpha\dot\alpha} -ig [A_{\alpha\dot\alpha},\ ]$
\begin{align}\label{Afield}
\mathcal{A}  
 =& \, 
A +
i \ket{ \theta^A} [ \bar\psi_{A}|
+
\frac{i}{2 !} \ket{\theta^A} \bra{\theta^B} D  \bar\phi_{AB} 
-
\frac{1}{3!} \epsilon_{ABCD} \ket{\theta^A} \bra{\theta^B} D \vev{ \theta^C   \psi^{D}} 
\\[3mm]
& \hfill +
\frac{i}{4!}  \epsilon_{ABCD} \ket{\theta^A} \bra{\theta^B} D \vev{ \theta^C | F | \theta^{D}}
+
\dots \, ,
\nonumber
\\[3mm]
\label{Ffield}
\mathcal{F}_A =& \, \frac{i}2 \bar\phi_{AB} \ket{\theta^B} 
-
\frac1{3!!}\epsilon_{ABCD}\ket{\theta^B} \vev{\theta^C \psi^D} 
+
\frac{i}{4!!} \epsilon_{ABCD}\ket{\theta^B}  \vev{\theta^C | F |  \theta^D}+\ldots
\, ,
\end{align}
where we used a compact notation for the quantities carrying spinor indices, e.g.
$\ket{\theta^A} \equiv \theta^A_\alpha$ and $[\bar\psi_A|\equiv \bar\psi_{\dot\alpha A}$ (see Appendix \ref{NotationsConvensions} for explanations). 
Notice that in the expression for $\mathcal{A}$, Eq.~\re{Afield}, 
the covariant derivatives act on the quantum fields only. The expansion of $\mathcal{A}$
starts with the gauge field and the additional factor $1/2$ in front of the bosonic connection in  
\re{Bsuperfield} is introduced to get the correct normalization for the lowest component, i.e., 
$\ft{1}{2} dx^{\dot\alpha\alpha} A_{\alpha\dot\alpha} = dx^\mu A_\mu$. The ellipses  in the 
right-hand sides of \re{Afield} and \re{Ffield}  denote higher-order terms in the $\theta$'s up to order 
$O(\theta^8)$. A characteristic feature of such terms 
is that, in the interaction-free limit (for $g = 0$),  they are proportional to the free equations of motion.%
\footnote{This is obvious from the supersymmetric transformations of the holomorphic gluon 
field strength
$
\delta_{Q} F^{\alpha\beta} =   
\left[ 
\epsilon^{\alpha A} (\bar\Omega_{\rm f})^{\beta}_{A} 
+  
\epsilon^{\beta A} (\bar\Omega_{\rm f})^{\alpha}_{A} \right]/4 +O(g)
$ with 
$
(\bar\Omega_{\rm f})^{\alpha}_{A} = \partial^\alpha{}_{\dot\alpha} \bar\psi^{\dot\alpha}_A
$.}
Thus, the  free-theory expansion of the bosonic and fermionic connections terminates at orders 
$O(\theta^4)$ and $O(\theta^3)$, respectively.

The form of the connections and the values of the rational coefficients in \re{Afield} and 
\re{Ffield} are fixed from the requirement for the supersymmetric Wilson loop \re{W-prod} 
to be invariant under the shift of the odd variables, $\delta \theta_\alpha^A=\epsilon_\alpha^A$, and the simultaneous chiral supersymmetric transformation of fields (see Eq.~\re{Q-susy} in Appendix \ref{NotationsConvensions}). It is a well-known feature of all supersymmetric gauge theories, considered in a non-supersymmetric gauge (Wess-Zumino gauge), that the supersymmetry algebra closes modulo compensating gauge transformations with a field-dependent parameter.  In addition, in the absence of auxiliary fields, as is the case of $\mathcal{N}=4$ SYM, the algebra closes on the shell of the field equations. An explicit calculation 
yields the following result for the $Q-$variation of both connections 
\begin{align}
\label{SUSYtransformAandF}
\delta_Q \mathcal{A}_{\alpha\dot\alpha} & = \partial_{\alpha\dot\alpha} \omega + i g [\omega, {\mathcal A}_{\alpha\dot\alpha}]
+ \Omega_{\alpha\dot\alpha}\,,\qquad
\\[2mm]\notag
\delta_Q{\mathcal F}_{\alpha A} & = \partial_{\alpha A}\omega + i g [\omega, {\mathcal F}_{\alpha A} ] 
\, ,
\end{align}
with $\omega$ being the field-dependent  gauge transformation parameter
\be
\label{Gaugeomega}
\omega 
=
\vev{\epsilon^{A} \theta^B}\left[
- \frac{i}{2!} \bar\phi_{AB}
+
\frac{1}{3!} \epsilon_{ABCD} \vev{\theta^C \psi^D}
-
\frac{i}{4!} 
\epsilon_{ABCD} \vev{\theta^C | F | \theta^D }
+
\dots
\right]
\, ,
\ee
and $\epsilon^A_\alpha$ being the parameter of the $Q-$transformations. Also, $ \Omega_{\alpha\dot\alpha}$ 
in the first relation in \re{SUSYtransformAandF} is given by
\ba
\label{EOMOmega}
\Omega_{\alpha\dot\alpha}
=
-
2
\epsilon_{ABCD}
(\epsilon^{\beta A} \theta_\beta^B) \theta^C_\alpha
\left[
\frac{1}{3!} (\Omega_{\rm f})_{\dot\alpha}^D
-
\frac{i}{4!} \theta_\gamma^D (\Omega_{\rm g})_{\dot\alpha}{}^\gamma
+ \dots
\right]
\, ,
\ea
and it 
is proportional to the fermion and gluon equations of motion
\be
\label{EOMoperators}
(\Omega_{\rm f})_{\dot\alpha}^A 
= D_{\dot\alpha}{}^\gamma \psi^A_\gamma
\, , \qquad
(\Omega_{\rm g})_{\dot\alpha}{}^\alpha
= D_{\dot\alpha}{}^\gamma F_\gamma{}^\alpha
\, .
\ee
Let us now examine the variation of the supersymmetric Wilson loop under the supersymmetric 
transformations \re{SUSYtransformAandF}.  Comparing the relations \re{SUSYtransformAandF} 
and \re{gauge}, we observe that the $\omega-$dependent terms in  \re{SUSYtransformAandF} 
can be eliminated in $\mathcal{W}_n$ by a compensating (infinitesimal) gauge transformation 
\re{gauge} with the parameter $U=1+ig \omega$. Thus, we are left over with the variation of the 
bosonic connection  proportional to the equations of motion, $\delta_Q \mathcal{A}_{\alpha\dot\alpha}
=\Omega_{\alpha\dot\alpha}$. Although such terms vanish on shell, that is for quantum fields 
satisfying their equations of motion, they do provide a nonvanishing contribution to the variation of
$\mathcal{W}_n$ under off-shell supersymmetry transformations. As we will see in a moment, 
this subtlety plays a very important r\^ole in testing the duality between supersymmetric Wilson 
loops and scattering amplitudes.

\subsection{Equivalent form of the supersymmetric Wilson loop}

As was explained in Sect.\ \ref{Matching}, to discuss the duality between supersymmetric
Wilson loops and superamplitudes it is convenient to switch 
from the $(x,\theta)-$coordinates to the momentum supertwistor $(\lambda,\mu,\chi)-$variables 
defined in \re{twistors}. Inverting the relations \re{twistors} we find with the help of
\re{super-mom}
\begin{align}\label{EtaToChi}
x_i^{\dot\alpha\alpha} 
= 
\frac{\mu_{i-1}^{\dot\alpha}\lambda_i^\alpha -\mu_{i}^{\dot\alpha}\lambda_{i-1}^\alpha }{\vev{i-1\, i}}
\,,\qquad
\theta_i^{\alpha A} = \frac{\chi_{i-1}^A \lambda_i^\alpha - \chi_i^A \lambda_{i-1}^\alpha}{
\langle i-1\, i \rangle}\,.
\end{align}
Rewriting the supersymmetric Wilson loop \re{W-prod} and \re{W-link} in terms of these variables, 
we obtain the equivalent form of $\mathcal{W}_n$ introduced in Ref.\ \cite{Car10}.
 
To begin with, we examine the expression for the Wilson line \re{W-link} and
replace the bosonic and fermionic connections in  \re{Bsuperfield} by their
explicit expressions \re{Afield} and \re{Ffield}. Then, we use the definition
\re{EtaToChi} to get, after a rather lengthy calculation, 
\be
\label{MStoCH}
 \mathcal{B}^{\rm MS} (t_j)
=
 \mathcal{E}_j (t_j)  
+ \lr{
\frac{d}{dt_j} \mathcal{V}_j (t_j) + ig [\mathcal{V}_j (t_j),  \mathcal{E}_j (t_j) ]}
+
\Delta \Omega_j (t_j)
\, ,
\ee
where the three terms in the right-hand side have the following meaning. The first term 
$ \mathcal{E}_j (t_j)$ depends on the single odd variable $\chi_j^A$ and its expansion 
runs to order $O(\chi_j^4)$ only
\ba
\label{CHexponentE}
2 \mathcal{E}_j (t)
&=&
- \langle j | A | j ]
-
i \chi_j^A [ \bar\psi_A | j ]
+\ft{i}{2!}
\chi_j^A \chi_j^B 
\frac{\langle j-1 | D | j ]}{\langle j-1 \,j \rangle} \bar\phi_{AB}
\nonumber\\
&-&
\ft{1}{3!}
\epsilon_{ABCD} \chi_j^A \chi_j^B \chi_j^C 
\frac{\langle j-1 | D | j ] \langle \psi^D | j - 1\rangle}{\langle j-1\, j \rangle^2}
\nonumber\\
&+&
\ft{i}{4!}
\epsilon_{ABCD} \chi_j^A \chi_j^B \chi_j^C \chi_j^D
\frac{\langle j-1 | D | j ] \langle j-1 | F | j-1 \rangle}{\langle j-1\, j \rangle^3}
\, ,
\ea
where we used the compact spinor notations explained in Appendix \ref{NotationsConvensions}.

The second term in the right-hand side of \re{MStoCH} involves the covariant derivative 
of the function
\ba\label{Vj}
\mathcal{V}_j
\!\!\!&=&\!\!\!
\ft{i}{2!}
\chi_j^A \frac{\langle j-1, \theta^B \rangle}{\langle j-1\, j \rangle} \bar\phi_{AB}
-
\ft{1}{3!}
\epsilon_{ABCD} \chi_j^A
\frac{\langle j-1, \theta^B \rangle}{\langle j-1\, j \rangle^2}
\langle \Lambda_j^C | \psi^D \rangle
\\ \nonumber 
&+&\!\!\!
\ft{i}{4!} \epsilon_{ABCD} \chi_j^A
\frac{\langle j-1, \theta^B \rangle}{\langle j-1\, j \rangle^3}
\Big\{
\langle \Lambda_j^C | F | \Lambda_j^D \rangle
+
\chi^C_j \langle j-1 | F | \Lambda_j^D \rangle
+
\chi^C_j \chi^D_j \langle j-1 | F | j-1 \rangle
\Big\} 
+ \dots \,,
\ea
where $\theta^A=\theta^A(t_j)$ was defined in \re{super-path} and 
a shorthand notation was introduced for the combination
\be
\ket{\Lambda_j^{C}} = \ket{j}  \langle j-1| \theta^C(t_j) \rangle - 2 \ket{j-1}  \chi_j^C
\, .
\ee
The contribution of $\mathcal{V}_j(t_j)$ to the Wilson line \re{W-link} 
can be factored out into boundary terms depending on $\mathcal{V}_j(t_j=1)$ and $\mathcal{V}_j(t_j=0)$. In the formulation 
of Ref.~\cite{Car10}, such terms get absorbed into the so-called 
vertex operators localized at the vertices of the super-polygon $\mathcal{C}_n$.

Finally, the last term in the right-hand side of  \re{MStoCH}  is proportional to the
equations of motion
\begin{align}\label{EOMinMSminusCH}
2\Delta \Omega_j 
=
- \ft{1}{3!}
\epsilon_{ABCD} \chi_j^A \chi_j^B
\frac{\langle j-1 \,\theta^C \rangle}{\langle j-1\, j \rangle}
[ \Omega_{\rm f}^{D}   j ]
-
\ft{i}{4!}
\epsilon_{ABCD} \chi_j^A \chi_j^B
\frac{\langle j-1 \,\theta^C \rangle}{\langle j-1\, j \rangle^2}
\langle \Lambda^C_j | \Omega_{\rm g} | j ]
\, ,
\end{align}
where $\Omega_{\rm f}$ and $\Omega_{\rm g}$ were defined in \re{EOMoperators}.
Neglecting the contribution of the latter to \re{MStoCH}, we can define another connection 
\begin{align}
\label{B-CH}
 \mathcal{B}^{\rm CH}(t_j)
=
 \mathcal{E}_j (t_j)  
+
\lr{
\frac{d}{dt_j} \mathcal{V}_j (t_j) + ig [\mathcal{V}_j (t_j),  \mathcal{E}_j (t_j) ]}
\, .
\end{align}
Then, we can use this function to construct the Wilson line \re{W-link} and
define the corresponding supersymmetric Wilson loop  \re{W-prod}. The resulting
expression for $\mathcal{W}_n^{\rm CH}$ coincides with the supersymmetric
Wilson loop introduced in Ref.~\cite{Car10}. However, since by construction
\begin{align}
\label{B-diff}
\mathcal{B}^{\rm MS}(t_j)-\mathcal{B}^{\rm CH}(t_j) = \Delta \Omega_j (t_j)\,,
\end{align}
the two Wilson loops proposed in Refs.~\cite{MasSki10,Car10} are identical only on shell. Thus, the difference between the all-loop expressions for the two Wilson loops can be
attributed to the  contribution of the field-equation operators.
 
As a consequence of our analysis, the natural question arises whether the equations of motion can 
produce a nonvanishing contribution to the expectation value of the supersymmetric Wilson 
loop \re{super-W}. If they do not, then the two Wilson loops $\mathcal{W}_n^{\rm MS}$ and 
$\mathcal{W}_n^{\rm CH}$ respect the $Q-$supersymmetry and they are identical at the quantum 
level. However, if the contribution of the equations of motion is different from zero, then 
$\mathcal{W}_n^{\rm MS}\neq \mathcal{W}_n^{\rm CH}$ and both Wilson loops are not $Q-$supersymmetry invariants. In the latter case, neither of the Wilson loops can be dual to the 
superamplitudes in $\mathcal{N}=4$ SYM since the latter  are invariant under the  
$Q-$supersymmetry.

\section{Supersymmetric Ward identities}
\label{SUSYWI}

To understand better what happens to supersymmetry at the quantum level,  let us derive the Ward identities 
for the two supersymmetric Wilson loops introduced in the previous section. 

We start with the path integral representation for the vacuum expectation value of the Wilson loop 
\be\label{path}
\langle \mathcal{W}_n(x_i,\theta_i) \rangle = \int [DX] \, \mathcal{W}_n(x_i,\theta_i)  \, {\rm e}^{i S_{\mathcal{N}=4}[X]}
\, ,
\ee
where the integration goes over all fields in $\mathcal{N}=4$ SYM collectively called $X$. Then, we make the shift  
$\theta_i\to \theta_i+\epsilon$ in both sides of \re{path} and perform a compensating supersymmetry transformation 
of the fields \re{Q-susy} inside the path integral, $X\to X+\delta_Q X$.
Making use of the invariance of the $\mathcal{N}=4$ action,  $S_{\mathcal{N}=4}[X]=S_{\mathcal{N}=4}[X+\delta X]$  
we arrive at %
\footnote{More precisely, the $Q-$variation of the gauge-fixed action is different from zero, but for the supersymmetry 
preserving regularization, i.e., dimensional reduction, $\delta_Q S_{\mathcal{N}=4}$ contains BRST-exact operators 
only and thus does not produce nontrivial
contributions when inserted into the correlation function with the gauge-invariant Wilson loop,
$\langle \mathcal{W}_n \left( i \delta_Q S_{\mathcal{N}=4} \right)\rangle = 0$ (see, e.g., \cite{BelMul00}). }
\begin{align}\label{Ward}
(\epsilon\cdot Q) \langle \mathcal{W}_n \rangle  \equiv
\sum_{i=1}^n \epsilon^A_\alpha \frac{\partial}{\partial \theta_{i \alpha}^A} \langle \mathcal{W}_n \rangle 
= 
\langle \delta_Q \mathcal{W}_n  \rangle
\, .
\end{align}
To make use of this relation we have to analyze the variation of
the supersymmetric Wilson loops $\delta_Q \mathcal{W}_n$, which we come to do next.

It follows from the definitions \re{W-prod} and \re{W-link} that the
supersymmetry variation of the Wilson loop $\mathcal{W}_n$ amounts to inserting a
 local operator $\delta \mathcal{B}(t_j)$ on the super-polygon $\mathcal{C}_n$
\begin{align}\notag
\vev{\delta_Q  \mathcal{W}_n} 
&= \sum_{j=1}^n  \frac1{N_c} \vev{ \tr  \lr{ \mathcal{W}_{[1,2]}  \ldots \delta_Q \mathcal{W}_{[j,j+1]} \ldots \mathcal{W}_{[n,1]}} }  
\\\label{delta-W}
&= \sum_{j=1}^n\, \frac{1}{N_c} \vev{\tr  \lr{ig \int_0^1 dt_j\, \delta_Q  \mathcal{B}(t_j)}  \mathcal{W}_n(t_i) }\,.
\end{align}
Here $\mathcal{W}_n(t_i)$ stands for the supersymmetric Wilson line
evaluated along an open contour in superspace that starts at the point 
$(x(t_i),\theta(t_i))$, goes along the polygon $\mathcal{C}_n$ and returns to the 
starting point.  

\subsection{Anomalies}

Let us first compute $\delta_Q  \mathcal{B}(t_j)$ for the supersymmetric Wilson loop 
$\mathcal{W}_n^{\rm MS}$ defined in \re{Bsuperfield}, \re{Afield} and \re{Ffield}. We apply 
the relation \re{SUSYtransformAandF} to get
\begin{align}\notag
\delta_Q  \mathcal{B}^{\rm MS}(t_j) 
&= \ft12 \dot x^{\dot\alpha\alpha}(t_j) 
\delta_Q {\mathcal A}_{\alpha\dot\alpha}+\dot \theta^{\alpha A}(t_j) \delta_Q{\mathcal F}_{\alpha A} 
\\ \label{B-con}
&=\frac{d \omega}{dt_j} +i g [\omega, \mathcal{B}^{\rm MS}(t_j)]
 -\ft12 \bra{j} \Omega(t_j) |j]\,,
\end{align} 
with $\omega$ and $\Omega$ defined in \re{Gaugeomega} and \re{EOMOmega}, respectively. As 
was explained in Sect.~\ref{SUSYconnection}, the terms involving $\omega$ do not contribute to the 
right-hand side of \re{delta-W} by virtue of gauge invariance. Therefore, computing \re{delta-W} we can 
retain only the last term in \re{B-con}
\begin{align}\label{Omega-MS}
\delta_Q  \mathcal{B}^{\rm MS}(t_j) = \Omega^{\rm MS}_j
= \epsilon_{ABCD}
\langle \epsilon^A \theta^B \rangle \vev{j\, \theta^C}
\bigg\{
\frac{1}{3!} [ \Omega_{\rm f}^D j]
+
\frac{i}{4!} \bra{\theta^D} \Omega_{\rm g}{}|j] + \dots
\bigg\}\,,
\end{align}
where $\theta^A=\theta^A(t_j)$ was defined in \re{super-path} and the
ellipses denote terms vanishing in the free theory limit  (i.e., for $g=0$).

The analysis of the Wilson loop $\mathcal{W}_n^{\rm CH}$ goes along the same lines.
We use the relations \re{CHexponentE} and \re{B-CH} to verify that, modulo gauge transformations,
\begin{align}\nonumber
\delta_Q  \mathcal{B}^{\rm CH}(t_j) = \Omega^{\rm CH}_j
& = 
\frac{1}{4} \epsilon_{ABCD} \frac{\chi_j^A \chi_j^B \langle j-1 \epsilon^C \rangle}{\langle j-1\, j\rangle}
[ \Omega_{\rm f}^{ D}  j ]
\\& \label{Omega-CH}
-
\frac{i}{12} \epsilon_{ABCD} 
\frac{\chi_j^A \chi_j^B \chi_j^C \langle j-1 \epsilon^D \rangle}{\langle j-1\, j\rangle^2}
\langle j-1 | \Omega_{\rm g} | j ] + \ldots
\, ,
\end{align}
where $\chi_j^A=\vev{j\theta_j^A}$. The obtained expressions for the anomalies have to satisfy a consistency condition that follows from \re{B-diff}
\begin{align}
\delta_Q  \mathcal{B}^{\rm MS}-\delta_Q  \mathcal{B}^{\rm CH}=\Omega^{\rm MS}_j-\Omega^{\rm CH}_j=\delta_Q  \Delta\Omega_j(t_j) \,.
\end{align}
Replacing $\Omega_j(t_j)$ by its explicit expression \re{EOMinMSminusCH} we find after lengthy 
calculations that this relation is indeed satisfied.

Substituting the obtained expressions for $\delta_Q \mathcal{B}$ into \re{delta-W} and \re{Ward} we 
deduce the Ward identity for the supersymmetric Wilson loop
\begin{align}
\label{susyWI}
(\epsilon \cdot Q) \langle \mathcal{W}_n \rangle  =  \sum_{j=1}^n\, ig
 \int_0^1 dt_j\, \VEV{\frac{1}{N_c}\tr \left[{   \Omega_j(t_j)}  \mathcal{W}_n(t_j)\right] }\,,
\end{align} 
where $\Omega_j(t_j)$ is given by \re{Omega-MS} and \re{Omega-CH} for the two supersymmetric 
Wilson loops defined above and the operator $(\epsilon\cdot Q)$ admits two equivalent forms
\begin{align}\label{Q-oper}
(\epsilon \cdot Q) = \sum_{j=1}^n \epsilon^A_\alpha \frac{\partial}{\partial \theta_{j \alpha}^A} 
= \sum_{j=1}^n \vev{i \epsilon^A} \frac{\partial}{\partial \chi_{j}^A} \,.
\end{align}
We would like to emphasize that that the right-hand side of \re{susyWI} involves 
the correlation function of the equations of motion with the Wilson loop.
For the Wilson loop to be invariant under the $Q-$supersymmetry, the right-hand side 
of \re{susyWI} should vanish. Otherwise, the $Q-$supersymmetry will be
broken.
  
\subsection{One-loop calculation}
 
Let us now compute the one-loop corrections to the right-hand sides of the Ward 
identities \re{susyWI} for both Wilson loops.  
We recall that the anomalies \re{Omega-MS} and \re{Omega-CH} are given by
a linear combination of the fermion and gluon equations of motion, $\Omega_{\rm f}$ 
and $\Omega_{\rm g},$ defined in \re{EOMoperators}.

We start with \re{Omega-MS} and 
retain for the moment only the terms involving $\Omega{}_{{\rm f}\, \dot\alpha}^A  = \partial_{\dot\alpha}{}^\gamma 
\psi^A_\gamma+O(g)$. Their contribution to the first term in the sum, i.e., for $j=1$, in the right-hand side of 
\re{susyWI} is given by (the remaining terms can be obtained through a cyclic shift of the indices)
\begin{align}\label{psi-B}
\frac{ig}{3!}\epsilon_{ABCD}\int_0^1dt_1
\langle \epsilon^A \theta^B(t_1) \rangle \vev{1 \theta^C(t_1)}
\VEV{\frac{1}{N_c}\tr \left\{[ \Omega_{\rm f}^D(x(t_1)) , 1] \lr{ig\int_{\mathcal{C}_n} dt\, \mathcal{B}(t)}\right\} }  +O(g^4)\,,
\end{align}
where we replaced the Wilson line $\mathcal{W}_n(t_1)$ by its lowest order expansion. Here the connection $\mathcal{B}(t)$ is
integrated over the contour $\mathcal{C}_n$ and the fermionic operator is inserted at the point 
$y=x(t_1)$ on the segment $[x_1,x_2]$. Computing the correlation function in  \re{psi-B} to the lowest order in the coupling, we have to Wick contract $\psi^A_\beta(x(t_1))$ 
from $\Omega_{\rm f}^D$ with the fermion field $\bar\psi_{\dot\beta B}(x(t))$ inside $\mathcal{B}(t)$. 
It is easy to see from \re{MStoCH}, \re{CHexponentE} and \re{B-CH} that the corresponding terms look alike for both 
connections and for the $k$th segment they are given by
\begin{align}\label{part-B}
\mathcal{B}(t_k) = -\frac{i}2  \chi_k^A\, [ \bar\psi_{A}(x(t_k))\,  k] +\ldots
\end{align}
In Eq.~\re{psi-B}, the $\bar\psi-$field can be located at any of 
the segments of the contour $\mathcal{C}_n$ including $[x_1,x_2]$. In that case, 
the two fields $\psi^A_\beta(x(t_1))$ and $\bar\psi_{\dot\beta B}(x(t_k))$ belong to the 
same segment and  they become light-like separated (recall that $x_{12}^2=0$).
Due to light-cone singularities, the product of two quantum fields separated by
a light-like interval is not well defined. Therefore, in order to
define the corresponding contribution to \re{psi-B} we have to introduce a regularization. 

In what follows we shall use the Four-Dimensional Helicity (FDH) regularization of $\mathcal{N}=4$ \cite{Bern:2002zk}.
The main advantage of this scheme is that it preserves the Poincar\'e supersymmetry of $\mathcal{N}=4$ SYM 
(at least to the lowest order in coupling as we do in our analysis) and allows
us to use the spinor decomposition for super-momenta \re{super-mom} without it interfering with the change of dimensionality
of Minkowski space-time. Then, the regularized correlator of the gaugino field with the fermion equation
of motion takes the form
\begin{align}\label{fermi-cor}
\frac1{N_c} \tr \langle \partial_{\dot\alpha}{}^\beta \psi^A_\beta(y) \bar\psi_{\dot\beta B} (x) \rangle 
= 
i {C_F} \delta^A_B \,
\frac{\varepsilon \Gamma (2 - \varepsilon)}{\pi^{2 - \varepsilon}} 
\frac{\epsilon_{\dot\alpha \dot\beta}}{[- (x-y)^2+i0]^{2 - \varepsilon}}
\, ,
\end{align}
where $C_F=(N_c^2-1)/(2N_c)$ is the quadratic Casimir of the $SU(N_c)$ gauge group.
According to the standard prescription for tadpoles \cite{Col82}, the correlation function \re{fermi-cor} vanishes 
for $(x-y)^2=0$ within the framework of dimensional regularization. This implies that
\re{psi-B} receives zero contribution when $\mathcal{B}(t)$ is integrated 
along the segment $[x_1,x_2]$.

Notice that the two-point function \re{fermi-cor} is proportional to the parameter
of dimensional regularization $\varepsilon$. Therefore, for the correlation function
in \re{psi-B} to be different from zero as $\varepsilon\to 0$ the integration over
the position of the fields in \re{psi-B} should produce a pole $1/\varepsilon$. 
As follows from \re{fermi-cor}, this could only happen if the two-point function 
\re{fermi-cor} is integrated through a
region where $(x-y)^2\to 0$. As an example, let us consider the contribution to 
\re{psi-B} when the connection $\mathcal{B}(t)$ is integrated along the segment $[x_2,x_3]$, that is for $k=2$ in \re{part-B}
\begin{align} \label{F-exam}
\text{Eq.\,\re{psi-B}} &= -
 \frac{1}{3!}\frac{g^2 C_F}{2\pi^{2 - \varepsilon}}\epsilon_{ABCD} \chi_1^C \chi_2^D [12] \int_0^1dt_1
     \int_0^1dt_2     
\frac{ \varepsilon \Gamma (2 - \varepsilon)\langle \epsilon^A \theta^B(t_1) \rangle}{[- x_{13}^2 (1-t_1) t_2+i0]^{2 - \varepsilon}}\,,
\end{align} 
where $x(t_i)=x_i-t_ix_{i,i+1}$, $\theta(t_i) = \theta_i-t_i\theta_{i,i+1}$ and we used of the identities 
$\vev{i\,\theta^A(t_i)} =\chi_i^A$ and
\be
(x(t_1)-x(t_2))^2= (x_{12}(1-t_1) + x_{23} t_2 )^2 = x_{13}^2 (1-t_1) t_2\,.
\ee
Here it is crucial that the two segments are light-like, $x_{i,i+1}^2=0$.
We observe that the integration in the right-hand side of \re{F-exam} around $t_2=0$, $t_1=1$ 
produces a pole $1/\varepsilon$. It compensates the factor of $\varepsilon$ coming from \re{fermi-cor}  
and produces a finite contribution
\begin{align}
 \text{Eq.\,\re{psi-B}}= 
\frac{1}{3!}\frac{g^2 C_F}{2\pi^2} \frac{[12]}{(x_{13}^2)^2}\epsilon_{ABCD} 
\langle \epsilon^A \theta_{12}^B \rangle \chi_1^C \chi_2^D
+ 
O(\varepsilon)\,.
\end{align}
 
This example illustrates the general mechanism which is at work for the fermion equations of motion and which produces a 
nonvanishing contribution to the right-hand side of \re{susyWI}.
Repeating the analysis for $k=3,\ldots,n$ in \re{part-B} it is straightforward to 
show that the right-hand side of \re{psi-B} only receives a nonzero contribution 
from four segments of the contour $\mathcal{C}_n$ adjacent to $[x_1,x_2]$,
that is from those with $k=2,3,n-1,n$.

The analysis of the contribution from the  gluon equations of motion to the right-hand 
side of \re{susyWI} goes along the same lines. To lowest order in the coupling, 
the gluon equation-of-motion operator  (\ref{EOMoperators}) entering  $\Omega_j(t_j)$ 
can only interact with the bosonic part of $\mathcal{W}_n$ given by 
$\ft{i}2\oint_{\mathcal{C}_n} dx^{\dot\alpha\alpha}
A_{\alpha\dot\alpha} (x)$. The correlation function of the gauge 
field with its equation-of-motion operator  (\ref{EOMoperators}) reads, in the Feynman 
gauge and to lowest order in the coupling,
\be\label{A-Omega}
\frac1{N_c} \tr \langle A_{\alpha\dot\alpha} (x_1) \Omega{}_{{\rm g} \, \dot\beta}{}^\beta (x_0) \rangle 
= i \frac{ C_F}{2 \pi^{2 - \varepsilon}} {\Gamma (2 - \varepsilon)} 
\bigg[
  (1 + \varepsilon)  
\frac{\epsilon_{\dot\alpha\dot\beta} \delta_\alpha^\beta}{(- x_{01}^2+i0)^{2 - \varepsilon}}
+
 {(1 - \varepsilon)} 
\frac{(x_{01})_{\dot\beta\alpha} (x_{01})_{\dot\alpha}{}^\beta}{(-x_{01}^2+i0)^{3 - \varepsilon}}\bigg]
\, .
\ee
Making use of this relation we find   the correlation function of the gluon equation 
of motion with the bosonic Wilson loop as
\ba
\frac1{N_c} \tr \label{GluonEOMpropagator}
\oint_{\mathcal{C}_n} dx^{\dot\alpha\alpha}
\langle A_{\alpha\dot\alpha} (x) \Omega{}_{{\rm g} \, \dot\beta}{}^\beta (x_0) \rangle 
&=&
i \frac{C_F}{ \pi^{2 - \varepsilon}}\,
 {\varepsilon \Gamma (1 - \varepsilon)}\sum_{j=1}^n
(x_{j,j+1})_{\dot\beta}{}^\beta D_\varepsilon (x_{j+1,0}, x_{j0})\,,
\ea
where the notation was introduced for
\be
D_\varepsilon (x_{j+1,0}, x_{j0}) 
= 
\frac{(- x_{j+1,0}^2+i0)^{\varepsilon - 1} - (- x_{j0}^2+i0)^{\varepsilon - 1}}{x_{j+1,0}^2 - x_{j0}^2}
\, .
\ee
In distinction with \re{A-Omega}, the relation \re{GluonEOMpropagator}
is gauge invariant.

Comparing \re{GluonEOMpropagator} with \re{fermi-cor} we observe the same
pattern. Though both correlation functions vanish for $\varepsilon\to 0$, they
produce a nonvanishing contribution to the right-hand side of \re{susyWI} upon 
integration over the polygon $\mathcal{C}_n$. We would like to emphasize that one may miss
this contribution if one  performs the calculation in $D=4$ dimensions without properly regularizing the light-cone singularities of 
the correlation functions.

\subsection{Supersymmetry anomalies}

It becomes straightforward to compute the one-loop correction to the right-hand side of \re{susyWI} 
for both Wilson loops, $\mathcal{W}_n^{\rm MS}$ and $\mathcal{W}_n^{\rm CH}$, by making use 
of the relations \re{fermi-cor} and  \re{GluonEOMpropagator}. For the sake of simplicity we present here the
explicit expressions for the simplest case of $n=4$, that is for the Wilson loop  defined over the rectangular 
contour $\mathcal{C}_4$ in the superspace. The generalization of our analysis to arbitrary $n$ is straightforward.

For the Wilson loop $ \mathcal{W}^{\rm CH}_4$ involving the superconnection \re{B-CH} the 
supersymmetric Ward identity reads
\begin{align}\label{CH-anom}
(\epsilon \cdot Q) \, \mathcal{W}^{\rm CH}_4  =
- \frac{g^2 C_F}{4 \pi^2} \epsilon_{ABCD}
\chi_1^A \chi_1^B \frac{\langle 4\epsilon^C \rangle}{\langle 41 \rangle}
\frac{[13]}{x_{13}^2 x_{24}^2}
\left( \chi_3^D + \frac{1}{3} \frac{\langle 34 \rangle}{\langle 41 \rangle} \chi_1^D \right)
+
{\rm (cyclic)} ,
\end{align}
where `(cyclic)' stands for terms obtained by the cyclic shift of indices $i\to i+1$ subject to the periodicity 
condition $i+4\equiv i$. Here the first and second terms inside the braces arise from the fermion 
and gauge equations of motion,  respectively.

For the Wilson loop $ \mathcal{W}^{\rm MS}_4$ involving the superconnection \re{Bsuperfield} the 
analogous supersymmetric Ward identity looks as
\begin{align}
(\epsilon \cdot Q)\, \mathcal{W}^{\rm MS}_4 
 & = 
\frac{g^2 C_F}{2 \pi^2} 
\frac{1}{3!} \epsilon_{ABCD}
\bigg\{
\langle \epsilon^A \theta_{12}^B \rangle \langle 1 \theta_1^C \rangle
\left(
\frac{[12]}{(x_{13}^2)^2} \langle 2 \theta_2^D \rangle
+
\frac{[41]}{(x_{24}^2)^2} \langle 4 \theta_4^D \rangle
\right)
\nonumber\\
& \qquad\qquad \qquad 
-
\frac{[13]}{x_{13}^2 x_{24}^2}
\left( \langle \epsilon^A \theta_1^B \rangle + \langle \epsilon^A \theta_2^B \rangle \right)
\langle 1 \theta_1^C \rangle \langle 3 \theta_3^D \rangle
\bigg\}
\nonumber\\
&- 
\frac{g^2 C_F}{2 \pi^2} 
\frac{1}{4!} \epsilon_{ABCD}
\bigg\{
\frac{[12]}{(x_{13}^2)^2}
\left(
\langle \epsilon^A \theta_{12}^B \rangle \langle 1 \theta_2^C \rangle \langle 2 \theta_2^D \rangle
+
\langle \epsilon^A \theta_2^B \rangle \langle 1 \theta_2^C \rangle \langle 2 \theta_{12}^D \rangle
\right)
\nonumber\\
& \qquad\qquad\qquad
-
\frac{[13]}{x_{13}^2 x_{24}^2}
\left( 
\langle \epsilon^A \theta_1^B \rangle \langle 1 \theta_1^C \rangle \langle 3 \theta_1^D \rangle
+
\langle \epsilon^A \theta_2^B \rangle \langle 1 \theta_2^C \rangle \langle 3 \theta_2^D \rangle
\right)
\nonumber\\ 
\label{MS-anom}
& \qquad\qquad\qquad
+
\frac{[41]}{(x_{24}^2)^2}
\left[ 
\langle \epsilon^A \theta_{12}^B \rangle \langle 1 \theta_1^C \rangle \langle 4 \theta_1^D \rangle
+
\langle \epsilon^A \theta_1^B \rangle \langle 1 \theta_1^C \rangle \langle 4 \theta_{12}^D \rangle
\right]
\bigg\}
+ 
{\rm (cyclic)}
\, .
\end{align}
Here the two terms in the right-hand side again describe the contribution of the fermion and gluon 
equations of motion, respectively.

The following comments are in order. 

The very fact that the right-hand sides of the relations \re{CH-anom} and 
\re{MS-anom} are different from zero immediately implies that  the chiral
$Q-$supersymmetry of both Wilson loops is broken already at one loop.
Moreover, it can be verified that the two anomalies do not respect the 
conformal symmetry (see Sect.\ \ref{SolvingWI} below).%
\footnote{In general, for some quantity depending on $\vev{ij}$ and $[ij]$ 
to be invariant under conformal transformations, the indices should satisfy the 
conditions $|i-j|=1$. This is not the case for the one-loop expressions for the anomalies, Eqs.~\re{CH-anom} 
and  \re{MS-anom}, which involve the square brackets $[13]$.} 
 
We recall that, if the supersymmetric Wilson loop $\mathcal{W}_4$ respected the
supersymmetry and conformal symmetry, it would be independent of the odd variables, 
Eq.~\re{super-W4}. In the next section, we will reconstruct the complete one-loop expression 
for the Wilson loops $\mathcal{W}_4$ by solving the Ward identities \re{CH-anom} and  
\re{MS-anom} and we will demonstrate that relation
\re{super-W4} is invalidated by the anomalies.
 
We established in this section, that the correlation functions of the fermion
and gluon equation of motion with the Wilson loop 
induce a nontrivial contribution to the supersymmetric Ward identity. Let us now show
that the same correlation functions allow us to compute  the difference
between the two Wilson loops under consideration, $\mathcal{W}_{4}^{\rm MS} -\mathcal{W}_{4}^{\rm CH}$.
Indeed, it follows from \re{B-diff} that to lowest order in the coupling
\begin{align}\label{W-W}
\mathcal{W}_{4}^{\rm MS} -\mathcal{W}_{4}^{\rm CH} = 
ig\int_0^1dt_1 \VEV{\frac1{N_c}\tr\left[  \Delta \Omega_1 (t_1) \lr{ig\int_{
\mathcal{C}_4} dt\, \mathcal{B}(t)}\right]} +\text{(cyclic)}\,,
\end{align}
where $\Delta \Omega_1 (t_1)$ is defined in \re{EOMinMSminusCH}. The latter is given 
by a linear combination of fermion and gluon equations of motion.
As a result, the calculation of the correlation function in the right-hand side
of \re{W-W}  can be easily performed along the same lines as in
\re{susyWI} and \re{psi-B}. Going through the derivation we find
\ba
\mathcal{W}_{4}^{\rm CH} 
\!\!\!&-&\!\!\!
\mathcal{W}_{4}^{\rm MS}
\nonumber\\
&=&\!\!\!
\frac{g^2 C_F}{4 \pi^2} \frac{1}{3!}
\epsilon_{ABCD} \chi_1^A \chi_1^B
\bigg\{
\frac{[13] \eta_1^C \chi_3^D}{x_{13}^2 x_{24}^2}
+
\frac{[12] \eta_1^C \chi_2^D}{x_{13}^4}
+
\frac{[41] \eta_1^C \chi_4^D}{x_{24}^4}
-
\frac{2 [13]}{x_{13}^2 x_{24}^2} \frac{\chi_3^D \chi_4^D}{\vev{41}}
\bigg\}
\nonumber\\
 &-&\!\!\!
\frac{g^2 C_F}{4 \pi^2} \frac{1}{4!} 
\epsilon_{ABCD} \chi_1^A \chi_1^B
\bigg\{
2 \frac{[12]}{x_{13}^4} \eta_1^C \chi_2^D  
+ 
2 \frac{[41]}{x_{24}^4} \eta_1^C \chi_4^D
+ 
\frac{[13]}{\vev{41}^2 x_{13}^2 x_{24}^2}
\bigg[
\chi_4^C \left( \vev{13} \chi_4^D + 2 \vev{34} \chi_1^D \right)
\nonumber\\ \label{W-W-1loop}
&+&\!\!\!
(\chi_4^C - \vev{41} \eta_1^C)
\left[ \vev{13} (\chi_4^D - \vev{41} \eta_1^D) + 2 \vev{34} \chi_1^D \right]
\bigg]
\bigg\}
+
{\rm (cyclic)}
\, , 
\ea
where $\eta_1$ was introduced in \re{super-mom}. It
can be re-expressed in terms of $\chi$'s using Eq.\ (\ref{EtaToChi}) as follows
\begin{align}
\eta^A_1
=
\frac{\chi_{4}^A}{\langle 41 \rangle}
+
\frac{\chi_{2}^A}{\langle 12 \rangle}
+
\frac{\langle 24 \rangle \chi_1^A}{\langle 41 \rangle \langle 12 \rangle}\,.
\end{align}
Applying the differential operator \re{Q-oper} to both sides of \re{W-W-1loop} we 
verified that $(\epsilon \cdot Q) (\mathcal{W}_{4}^{\rm CH} -\mathcal{W}_{4}^{\rm MS})$ 
coincides with the difference of the expressions in the right-hand sides of Eqs.\ \re{CH-anom} and \re{MS-anom}.

\section{Supersymmetric Wilson loops at one loop}
\label{OneLoopWL}

In this section, we will compute one-loop correction  to the supersymmetric Wilson loop $\mathcal{W}_4$.
As follows from its definition \re{W-prod} and \re{W-link},  the Wilson loop is given to this order by the 
following expression
\begin{align}\label{W4-double}
\mathcal{W}_4 
= 1+ \sum_{1\le j\le k \le 4}(ig)^2\int_0^1 dt_j \int_0^1 dt_k\,\frac1{N_c}\tr \VEV{\mathcal{B}_j(t_j) \mathcal{B}_k(t_k)} 
+O(g^4)\,,
\end{align}
where the superconnection $\mathcal{B}$ is defined in \re{MStoCH} and \re{B-CH}.
The explicit expression for the superconnection involves the sum over all quantum fields in $\mathcal{N}=4$ SYM.  
As a consequence, the direct calculation of \re{W4-double} yields a large number of contributing terms 
and makes the analysis very cumbersome. There is however another method that allows one to efficiently fix 
the form of the one-loop Wilson loop with little effort. Namely,  we will solve the supersymmetric Ward identities, 
Eqs.~\re{CH-anom} and \re{MS-anom}, and reconstruct the explicit one-loop expression for the Wilson 
loop $\mathcal{W}_4$ using some input requiring only a very small number of diagrams being computed 
explicitly.

We recall that the 
expansion of  $\mathcal{W}_4$ in powers of the odd variables has the general form \re{W-exp}. To one-loop 
order, this expansion terminates at %
\footnote{To see this we notice that the one-loop correction to $\mathcal{W}_4$ 
in \re{W4-double}  is bilinear 
in the superconnecton $\mathcal{B}(t)$ which in turn is a polynomial of degree 4 in the Grassmann variables.}
\begin{align}\label{W4-ansatz}
\mathcal{W}_4 = \mathcal{W}_{4;0} + \mathcal{W}_{4;1} + \mathcal{W}_{4;2} + O(g^4)\,,
\end{align}
where $\mathcal{W}_{4;0}$ is the bosonic light-like Wilson loop and $\mathcal{W}_{4;k=1,2}$ 
are given by homogenous polynomials of degree $4k$ in the $\chi$'s. 
According to \re{CH-anom} and \re{MS-anom}, the anomaly $(\epsilon\cdot Q) \mathcal{W}_4$ 
is given to one-loop order by a homogenous polynomial in the odd variables
of degree 4. Then, replacing $\mathcal{W}_4$ by its general expression \re{W4-ansatz} and 
matching the degree of the odd variables we find that $\mathcal{W}_{4;0}$
and $\mathcal{W}_{4;2}$ should be annihilated by  $(\epsilon\cdot Q)$. For the 
bosonic component $\mathcal{W}_{4;0}$ this is obvious, while for $\mathcal{W}_
{4;2}$ it leads to a nontrivial constraint $(\epsilon\cdot Q)\, \mathcal{W}_{4;2} =O
(g^4)$. As we will see in a moment,  $\mathcal{W}_{4;2}$ takes zero value at one loop.
Thus, to this order, the supersymmetric Ward identities become anomalous for 
the component $\mathcal{W}_{4;1}$ only. By definition, $\mathcal{W}_{4;1}$ is a homogenous 
polynomial of degree 4 in $\chi_i$ (with $i=1,\ldots,4$). To one-loop
order, we shall use the following ansatz  for $\mathcal{W}_{4;1}$
\footnote{Here, to simplify the notation, we stripped the 
$SU(4)$ indices off the Grassmann variables and the accompanying Levi-Civita tensor these
are contracted with. As explained in Appendix \ref{NotationsConvensions}, in this form the $\chi$'s 
can be treated as commuting variables.} 
\begin{align}\label{W4-c}
\mathcal{W}_{4;1} = \frac{g^2 C_F}{4 \pi^2}  \sum_{0\le k_i \le 4
\atop
k_1+k_2+k_3+k_4=4} c_{k_1k_2k_3k_4} (\chi_1)^{k_1} (\chi_2)^{k_2} (\chi_3)^{k_3} (\chi_4)^{k_4}   
\, ,
\end{align}
where $c_{k_1k_2k_3k_4}$ are bosonic coefficient functions. As follows from Eq.\ \re{W-prod},
the Wilson loop $\mathcal{W}_4$ is invariant under the cyclic shift of indices, thus we have to require 
that the right-hand side of \re{W4-c} should be cyclically symmetric as well. 

The general solution to the Ward identities  \re{CH-anom} and \re{MS-anom}
is defined up to an arbitrary function depending on the invariants 
of the $Q-$supersymmetry transformations, $\chi_i^A \to \chi_i^A + \vev{i \epsilon^A}$. 
Such invariants depend on three points and have the following form
\footnote{To construct these invariants we use the transformations 
$\chi_i^A \to \chi_i^A + \vev{i \epsilon^A}$ and choose $\epsilon^A$ to put to
zero two of the $\chi$'s, e.g. $\chi_j=\chi_k=0$. Then, the remaining $\chi-$variables will be automatically
invariant under $Q-$supersymmetry.}
\begin{align}
\Theta_{ijk}^A = \chi_i^A \vev{jk} + \chi_j^A \vev{ki} + \chi_k^A \vev{ik}\,.
\end{align}
For $n$ points there are $(n-2)$ linear independent invariants. For $n=4$ we can choose 
them to be $\Theta_{412}^A$ and $\Theta_{234}^A$.
Then, the solution to the Ward identities are defined modulo the substitution
\begin{align}\label{W4-amb}
\mathcal{W}_4 \to  \mathcal{W}_4 + f_0(x) + f_1(x;\Theta_{412},\Theta_{234}) +
f_2(x) (\Theta_{412})^4 (\Theta_{234})^4\,,
\end{align}
where $f_0(x)$ and $f_2(x)$ are arbitrary functions of the bosonic variables
and $f_1(x;\Theta_{412},\Theta_{234})$ is an arbitrary homogenous polynomial in the odd 
variables of degree 4.   

\subsection{Boundary conditions}

To define a unique solution to the Ward identities, we have to fix the ambiguity 
in \re{W4-amb}, or equivalently determine the functions $f_0$, $f_1$ and $f_2$.
The function $f_0(x)$ affects only the bosonic component and, therefore, its form
is fixed by the one-loop correction to the bosonic Wilson loop $W_4$.
To determine the remaining functions $f_1$ and $f_2$, we shall compute one-loop 
corrections to $\mathcal{W}_4$ in the gauge $\chi_2^A=\chi_4^A=0$. In this gauge,
two major simplifications occur. Firstly, the expression for the superconnections
significantly reduce (see, e.g., Eqs.~\re{CHexponentE} and \re{Vj}),  thus minimizing the 
number of one-loop Feynman diagrams contributing to the loop. Secondly, the 
$Q-$invariants now depend on a single odd variable, $\Theta_{412}=\vev{24}\chi_1$  and 
$\Theta_{234}=-\vev{24}\chi_3$, and their contribution to \re{W4-amb} takes a particularly simple form.

Let us start with the last term in the right-hand side of \re{W4-amb}. In the gauge
$\chi_2=\chi_4=0$, it is proportional to $\eta_1^4\eta_3^4$. Using the expression for 
one-loop corrections to the Wilson loop \re{W4-double}, we find that $O(\eta_1^4\eta_3^4)$ 
term could only appear from the correlation between $O(\eta_1^4)$ 
and  $O(\eta_3^4)$ terms inside $\mathcal{B}_1(t_1)$ and $\mathcal{B}_3(t_3)$. The 
corresponding correlation function involves 
chiral components of the gauge field strength tensor located at two different segments of
the polygon, $\vev{F_{\alpha\beta}(x(t_1))F_{\alpha'\beta'}(x(t_3))}$. In $D=4$ dimension it vanishes, 
while for $D=4-2\varepsilon$ it is proportional to $\varepsilon$. Therefore, for the result to be different 
from zero, the integral over $t_1$ and $t_3$ should produce a pole $1/\varepsilon$. A simple calculation 
shows that this does not happen and, therefore, the coefficient in front of $O(\eta_1^4\eta_3^4)$ term in 
$\mathcal{W}_4$ equals zero.  This immediately implies that the one-loop correction to $\mathcal{W}_4$ 
does not involve terms of degree 8 in odd variables,
\be\label{zero}
\mathcal{W}_{4;2}=0+O(g^4)\,.
\ee

We now turn to computing corrections to \re{W4-double} of the Grassmann degree 4.
They have the general form \re{W4-c}. For  $\chi_2 = \chi_4 = 0$ we are left with
the terms of the form $\chi_1^{j_1}\chi_3^{4-j_1}$ with $j_1=0,\ldots,4$. As we
will show in the next subsection, to construct a unique solution to the Ward
identity it is sufficient to identify the contribution of two terms only, 
$O(\chi_1^2\chi_3^2)$ and $O(\chi_1\chi_3^3)$. Using explicit expressions for
the superconnections, Eqs.~\re{Afield} and \re{Ffield}, it is easy to see that 
such terms arise from the correlation between scalar and fermion fields, respectively, 
entering into the expansion of $\mathcal{B}(t_1)$ and $\mathcal{B}(t_3)$.

Since the difference between the two superconnections, Eqs.~\re{B-diff} and \re{EOMinMSminusCH}, 
does not involve scalar operators, the contribution of $O(\chi_1^2\chi_3^2)$ terms to the two loops, 
$\mathcal{W}_4^{\rm MS}$ and $\mathcal{W}_4^{\rm CH}$, is the same. 
To compute it we retain in the right-hand side of \re{W4-double}
only one term with $j=1$, $k=3$ and replace
\begin{align}
\mathcal{B}(t) =- \ft{i}{4}   \dot x^{\dot\alpha\alpha} 
\theta_\alpha^A \theta_\beta^B D^\beta{}_{\dot\alpha} \bar\phi_{AB}  
+
\ft{i}{2} 
 \dot\theta^{\alpha A} \theta_\alpha^B\,\bar{\phi}_{AB} +\ldots
\end{align}
A simple calculation shows that the coefficient in front of $\chi_1^2 \chi_3^2$  vanishes, or in 
application to \re{W4-c}   $c_{2020} = 0$. Then, the cyclic invariance of the Wilson loop \re{W4-c} 
implies that 
\begin{align}\label{bc1}
c_{2020} = c_{0202} = 0\,.
\end{align}
We remind that these relations hold for both Wilson loops. 

Let us now examine $O(\chi_1\chi_3^3)$ contribution to \re{W4-double}. By virtue of  \re{B-diff}, it takes 
a different form for the two Wilson loops. For $\mathcal{W}_4^{\rm MS}$, 
we replace the superconnection in \re{W4-double} by the following expressions (see Eqs.~\re{Afield} and \re{Ffield})
\begin{align}\notag
\mathcal{B}^{\rm MS}(t_1) 
&= 
\ft{i}{2} \langle 1 \theta^A_1 \rangle   [1 | \bar\psi_A (x_1 - t_1 x_{12}) ]+\ldots
\, , \\[2mm]
\mathcal{B}^{\rm MS}(t_3) 
&= 
\ft{1}{12} \epsilon_{ABCD}  
\left(
\langle 3 \theta^A_3 \rangle
[ 3 | \partial | \theta^B \rangle
+
4 \langle \theta_3^A \theta_{4}^B \rangle
\right)
\langle \theta^C | \psi^D (x_3 - t_3 x_{34}) \rangle+\ldots
\, ,
\end{align}
and find that the coefficient in front of $\chi_1\chi_3^3$ vanishes,  
$c^{\rm MS}_{1030}=0$. Again, the cyclic invariance of  \re{W4-c} leads to
\begin{align}\label{bc2}
c^{\rm MS}_{1030}=c^{\rm MS}_{0103}=c^{\rm MS}_{3010}=c^{\rm MS}_{0301}=0\,.
\end{align}
For the Wilson loop $\mathcal{W}_4^{\rm CH}$, we use the fact that the difference
between the superconnections \re{B-diff} is proportional to the fermion equations
to motion to obtain
\begin{align}\label{c-CH}
c^{\rm CH}_{1030}=
-\frac{1}{6}   
\frac{\langle 24 \rangle^2}{\langle 23 \rangle^2 \langle 34 \rangle \langle 41 \rangle x_{13}^2}
=\frac{1}{6}  \lr{\frac1{x_{13}^2}+\frac1{x_{24}^2}}\frac{\vev{24}}{\vev{23}\vev{34}\vev{13}}
\,,
\end{align}
where in the last relation we made use of the identities  
\begin{align}
\frac{x_{13}^2}{x_{24}^2} = -\frac{\vev{12}\vev{34}}{\vev{23}\vev{41}}\,,\qquad 
\frac{x_{13}^2+x_{24}^2}{x_{24}^2} =-\frac{\vev{13}\vev{24}}{\vev{23}\vev{41}}\,.
\end{align}
The same result \re{c-CH} can be deduced from \re{W-W-1loop} by examining the coefficient in front of 
$\chi_1\chi_3^3$ in $\mathcal{W}_{4}^{\rm MS} -\mathcal{W}_{4}^{\rm CH}$. 

\subsection{Solving the Ward identities}
\label{SolvingWI}

We are now in a position to construct the solution to the Ward identities \re{CH-anom} and \re{MS-anom}. 
We start with the former and substitute the ansatz \re{W4-c} into the left-hand side of \re{CH-anom}. 
In this way we obtain a relation both sides of which depend on the parameters of the supersymmetric 
transformation $\vev{i\epsilon^A}$ (with $i=1,\ldots,4$). Notice that among the four parameters only 
two are linearly independent 
\begin{align}
\vev{2 \epsilon} =  \vev{1 \epsilon}\frac{\vev{23}}{\vev{13}} + \vev{3 \epsilon}\frac{\vev{12}}{\vev{13}}
\, ,
\qquad
\vev{4 \epsilon} =  - \vev{1 \epsilon}\frac{\vev{34}}{\vev{13}}- \vev{3 \epsilon}\frac{\vev{41}}{\vev{13}}\,.
\end{align}
Then, comparing the coefficients in front of various terms involving $\vev{1\epsilon}$, $\vev{3\epsilon}$ 
and different powers of $\chi_i$, we obtain a system of linear inhomogeneous equations for the coefficients 
$c_{k_1k_2k_3k_4}$ entering  \re{W4-c}. In addition, we impose the boundary conditions \re{bc1} and \re{bc2}. 
The resulting system of equations is overdetermined but it has a unique solution leading to the 
following expression for the one-loop Wilson loop  
\begin{align}\notag
\mathcal{W}_{4;1}^{\rm MS} 
&=
\frac{g^2 C_F}{4\pi^2} \left( \frac{1}{x_{13}^2} + \frac{1}{x_{24}^2} \right) 
\bigg\{
- \frac{\vev{24}^2}{\vev{12}^2\vev{41}^2}  \frac{\chi_1^4}{24}
-
\left( 
\frac{\vev{24}}{\vev{12} \vev{41}^2} \chi_4
+
\frac{\vev{24}}{\vev{12}^2 \vev{41}} \chi_2
\right) \frac{\chi_1^3}{6}
\\\label{MSsolutionWI}
&
+
\bigg( 
\frac{\chi_2 \chi_3}{\vev{12} \vev{13}}
 - 
\frac54 \frac{\chi_2 \chi_4}{\vev{41} \vev{13}}
-
\frac{\chi_3 \chi_4}{\vev{41} \vev{13}}
-
\frac{1}{4}
\frac{3 \vev{24} \vev{13} + 2 \vev{34} \vev{12}}{\vev{12}^2 \vev{13} \vev{24}} \chi_2^2
\bigg) \frac{\chi_1^2}{3}\bigg\}
+ {\rm (cyclic)}
\, .
\end{align}
In a similar fashion, solving the Ward identity \re{CH-anom} subject to 
the boundary condition  \re{bc1} and \re{c-CH} we find
\begin{align}
\label{CHsolutionWI}
\mathcal{W}_{4;1}^{\rm CH} 
&= 
\frac{g^2 C_F}{4 \pi^2} \left( \frac{1}{x_{13}^2} + \frac{1}{x_{24}^2} \right) 
\bigg\{
\frac{\vev{23}\vev{24}}{\vev{12}^2\vev{13}\vev{41}} 
\frac{\chi_1^4}{12}
\nonumber\\
&+ 
\bigg(
\frac{\vev{24}}{\vev{41}\vev{12}\vev{13}} \chi_3
-
\frac{\vev{13}\vev{24} + \vev{12}\vev{34}}{\vev{12}\vev{13}\vev{41}^2} \chi_4
+
\frac{\vev{23}\vev{41} - \vev{13}\vev{24}}{\vev{12}^2\vev{13}\vev{41}} \chi_2  
\bigg) 
\frac{\chi_1^3}{6} 
\nonumber\\
&+ 
\bigg(
\frac{\chi_2\chi_3}{\vev{12}\vev{13}}
-
\frac{\chi_3\chi_4}{\vev{13}\vev{41}} 
- 
\frac{\chi_2\chi_4}{\vev{12}\vev{41}}
-
\frac{\chi_2^2}{\vev{12}^2}
\bigg) \frac{\chi_1^2}{2}
\bigg\} 
+ {\rm (cyclic)}
\, .
\end{align}

We verified that these relations obey several consistency conditions. By construction, they satisfy 
the Ward identities \re{CH-anom} and \re{MS-anom}, respectively, and,
therefore, they do not respect $Q-$supersymmetry
\begin{align}\label{Q-W}
Q^A_\alpha\, \mathcal{W}_{4;1}\neq 0\,.
\end{align} 
Also, the difference between the 
two Wilson loops is in agreement with \re{W-W-1loop}. 

Next, we computed $O(\chi_1\chi_2\chi_3\chi_4)$ 
and $O(\chi_1^4)$ terms in the right-hand side of  \re{W4-double} and checked that they are correctly 
reproduced in \re{MSsolutionWI} and \re{CHsolutionWI}. In the first case, the terms involving all four 
$\chi-$variables cannot be produced by the correlation function of two superconnections (see 
\re{W4-double}) and, therefore, they should be absent in the one-loop approximation. In the second 
case, to identify the terms $O(\chi_1^4)$ we can set $\chi_2=\chi_3=\chi_4=0$ thus reducing significantly 
the number of contributing diagrams.

We recall that the  relations  \re{MSsolutionWI} and \re{CHsolutionWI} define (together with
\re{W4-ansatz}) the part of the one-loop correction to the Wilson loop \re{W4-ansatz} depending on 
the Grassmann variables.  The bosonic component of the Wilson loop, $\mathcal{W}_{4;0}$, has 
been previously computed in Ref.~\cite{KorDruSok08}. 
To one-loop order, it develops a double pole $1/\varepsilon^2$ due to the presence of 
specific ultraviolet (UV) cusp singularities \cite{KK92}. 
In distinction with  $\mathcal{W}_{4;0}$, the expressions in the right-hand side of \re{CHsolutionWI} 
and \re{MSsolutionWI} are free from any divergences and are well-defined for $\varepsilon\to 0$. 
This implies that UV divergences cancel in the ratio of the supersymmetric Wilson loop and 
its bosonic component, 
\begin{align}\label{W/W}
\mathcal{W}_4/\mathcal{W}_{4;0}=1+\mathcal{W}_{4;1} 
+O(g^4)\,,
\end{align}
 and, 
therefore, the scaling (dilatation) invariance is restored in their ratio, $D\,\lr{\mathcal{W}_4/\mathcal{W}_{4;0}} 
=0$. But what about  special conformal transformation?

The special conformal boosts are given by the superposition of translations
and inversions, $K^{\dot\alpha\alpha}=I  P^{\dot\alpha\alpha}  I$. Translations shift $x_i$ but they do not affect
$\lambda_i$ and $\chi_i$. At the same time, the inversion acts as  
\begin{align}\label{Inv}
I[(x_i)_{\alpha\dot\alpha}] = \frac{x_i^{\dot\alpha\alpha}}{x_i^2}\,,\qquad
I[\lambda_i^\alpha] = (\mu_i)_{\dot\alpha}\,,\qquad I[\chi_i] =\chi_i\,,
\end{align} 
with $\mu_i$ defined in \re{twistors}.
A close examination of \re{CHsolutionWI} and \re{MSsolutionWI} shows  that both expressions depend 
on the angle brackets of the form $\vev{i\,i+1}$ and $\vev{i\,i+2}$. Using \re{Inv} we find that the former 
brackets are transformed covariantly under the inversion, 
$I[\vev{i\, i+1}] = \vev{i|x_i x_{i+1}|i+1} = x_{i+1}^2 \vev{i\,i+1}$, but this is not the case for the brackets 
of the type $\vev{i\,i+2}$. It is then straightforward to verify that both expressions are transformed nontrivially 
under inversion and, therefore, do not respect the special conformal invariance
\begin{align}\label{K-W}
K^{\dot\alpha\alpha} \mathcal{W}_{4;1}\neq 0\,.
\end{align}
This conformal anomaly is explicitly calculated in Appendix \ref{conf-anom}, where we show complete consistency 
with the solution of the supersymmetric Ward identity that we constructed in Sect. \ref{SUSYWI}.

Combining together the relations \re{Q-W} and \re{K-W} we conclude that the
ratio of the Wilson loops in \re{W/W} does possess neither $Q-$supersymmetry,
nor conformal (super)symmetry already at one loop. As was already explained in Introduction, 
this fact is in contradiction with the relations \re{W-sym} and \re{KW=0} which follow  in 
their turn from the conjectured duality between supersymmetric Wilson loops and scattering 
amplitudes in $\mathcal{N}=4$ SYM.

\section{Conclusions}
\label{Conclusions}

The MHV scattering amplitudes in planar $\mathcal{N}=4$ SYM are dual to bosonic Wilson loops. In 
this paper, we explored two recent proposals for extending this duality to generic non-MHV amplitudes. The 
corresponding dual object should have the same symmetries as the scattering amplitudes and be invariant 
to all loops under the chiral half ($Q-$ and $\bar S-$symmetries) of the  $\mathcal{N}=4$ superconformal 
symmetry. The supersymmetric extensions of the bosonic Wilson loop proposed in Ref.~\cite{Car10}
comply with this condition at the classical level but only up to terms proportional to field equations. We would like to  
point out that while our conclusions apply to  the supersymmetric Wilson loop in Minkowski space, the twistor space
version from Ref.~\cite{MasSki10} appears to be formulated off-shell and thus the question of whether any pathologies 
permeate this formalism as well requires extra studies.
 
Examining the properties of the supersymmetric Wilson loops at the quantum level,
one encounters the following complication. According to its definition, the perturbative expansion
of the Wilson loop involves products of various fields (scalars, gauginos and gluon) 
integrated along a closed contour. Since the integration contour is uniquely determined by 
the light-like momenta of the scattered particles, these fields inevitably become light-like 
separated. In the classical theory such an object is well defined in $D=4$ dimensions, while at 
the quantum level it suffers from light-cone singularities and requires regularization.
This subtlety is not a specific feature of the supersymmetric Wilson loop and it is already 
present in the bosonic light-like Wilson loop. In the latter case, due to the light-cone 
singularities of the two-point functions of the gauge fields, the quantum corrections to the bosonic 
Wilson loop generate ultraviolet divergences which appear  as double poles in $\varepsilon$ in the dimensional 
regularization scheme with $D=4-2\varepsilon$. By the same
token, the calculation of loop corrections to the supersymmetric Wilson loop demands 
the use of a regularization which we chose to be the supersymmetry preserving 
dimensional reduction.%
\footnote{It is important to point out that there is a difference in the use of dimensional reduction 
and dimensional regularization already at one loop. Had we used dimensional regularization 
instead, the supersymmetry Ward identities would not be fulfilled indicating inconsistencies in the
treatment at the quantum level.}
We would like to emphasize that even though the final expressions for the one-loop correction to 
$\mathcal{W}_{4;1}$ (see Eqs.~\re{CHsolutionWI} and \re{MSsolutionWI}) are finite as $\varepsilon\to 0$, 
they arise by adding
together the contributions from several Feynman diagrams, each of which develops light-cone 
divergences and, therefore, requires a regularization.%

Since the calculation of the quantum corrections to the supersymmetric Wilson loop involves
going away from the critical dimension $D=4$, one might suspect that some of the
classical symmetries will be broken at the quantum level. Indeed,
we demonstrated in this paper that, for the supersymmetric Wilson loops under consideration,  
both the chiral supersymmetry and conformal invariance are already broken at one loop. 
The underlying mechanism looks as follows. The variation of the Wilson loop under
supersymmetry $Q-$transformations is proportional to a contribution from the equation of 
motion operators. In dimensional regularization with $D=4-2\varepsilon$, the latter scales as 
$O(\varepsilon)$ and vanishes when the number of space-time dimensions is set to four. 
However, the correlation function of the equations of motion with the light-like Wilson loop 
produces a divergent $O(1/\varepsilon)$ contribution.\footnote{This effect is reminiscent to the 
so-called $\mu_\varepsilon^2-$terms in the scattering amplitudes. The latter appear as $O(\varepsilon)$ 
terms in the expression for the {\it integrand} of the dimensionally regularized amplitudes, but they 
produce a nontrivial contribution after integration over loop momenta starting from two loops \cite{BerDixKosRoiSprVerVol08,KosRoiVer10}.} 
Combining together the two effects, we obtain a nontrivial finite $O(\varepsilon^0)$ contribution to 
the corresponding supersymmetric Ward identity. The latter takes the form of a differential equation  
in the Grassmann variables and completely fixes the form of the super-Wilson loops with a minimal 
perturbative input. The unique solutions that we  found were tested against explicit one-loop calculations of 
the Wilson loops. 

Our analysis explicitly demonstrates that the quantum anomalies break the chiral 
supersymmetry of the Wilson loops, thus invalidating the conjectured duality with
the scattering amplitudes in $\mathcal{N}=4$ SYM. We recall that for the amplitudes
the same symmetry (the dual chiral $Q-$supersymmetry) remains exact at the quantum level. In 
fact, it is a trivial consequence of the way the dual variables are introduced in Eq.~\re{super-mom}. 
Notice that the general solution to the anomalous supersymmetriy Ward identity is defined
up to an arbitrary supersymmetric invariant function which could be compared with the 
scattering amplitude. In application to the Wilson loop
this would mean that while the super Wilson loop is affected by nonvanishing contributions of the  
equations of motion, it is its supersymmetric invariant part with the properly subtracted 
anomaly which is matched with the scattering amplitudes. This would require, however, a detailed knowledge 
of the all-loop anomaly as well as a consistent formulation of a scheme for separation between 
anomalous and invariant terms inside the Wilson loop.

The fact that the anomaly in the $Q-$supersymmetry is due to the contributions of the field equations to the Wilson loop correlation function is closely related to the necessity to work with an on-shell realization of the $\mathcal{N}=4$ supersymmetry algebra. The problem we encounter can be presented as follows. The bosonic superconnection $\mathcal{A}_{\alpha\dot\alpha}(x,\theta)$ introduced in \re{Afield}, can be viewed as the supersymmetric extension of the gauge field $A_{\alpha\dot\alpha}(x)$ obtained by  a finite chiral supersymmetry transformation with generators $Q^\alpha_A$ and with the Grassmann variables $\theta^A_\alpha$ in the role of the  parameter:
\begin{align}\label{expo}
\mathcal{A}_{\alpha\dot\alpha}(x,\theta) = e^{(\theta\cdot Q)}  A_{\alpha\dot\alpha}(x)\,,
\end{align}
with $(\theta\cdot Q)=\theta^A_\alpha Q^\alpha_A$.
In expanding the exponential we make use of the relation $\{ Q^\alpha_A, Q^\beta_B\}=0$ which follows from the chiral supersymmetry algebra. However, in the $\mathcal{N}=4$  case even this chiral subalgebra of  the full Poincar\'e supersymmetry closes only on shell. Then it is clear that the bosonic connection constructed from (\ref{expo}) is invariant under $Q-$transformations (up to a compensating gauge transformation) only modulo field equations. 

In supersymmetric theories the well-known remedy consists in adding sets of so-called auxiliary fileds, whose equations of motion are algebraic and whose role is to maintain the closure of the supersymmetry algebra off shell. If we had the relevant auxiliary fields at hand, our super Wilson loop would presumably not suffer from the anomaly we observed. An example, to be presented elsewhere, is provided by $\mathcal{N}=1$ SYM. In this simplest supersymmetric gauge theory one needs to add just a single scalar auxiliary field in order to close the algebra off shell. The $U(1)$ R symmetry of $\mathcal{N}=1$ SYM does not allow us to construct a purely chiral extension of the Wilson loop. Instead, we may consider the fully supersymmetric $\mathcal{N}=1$ super Wilson loop, with and without the auxiliary field in it. This will give us an alternative view on the anomaly mechanism. However, coming back to the  $\mathcal{N}=4$  case, we have to recall the very old result of \cite{Siegel:1981dx} on the absence of  auxiliary fields for  $\mathcal{N}=4$  SYM. In view of this, the  $\mathcal{N}=4$ super Wilson loop seems condemned to suffer from supersymmetry anomalies. The only escape might be that the no-go argument of \cite{Siegel:1981dx} applies to the full  $\mathcal{N}=4$  supersymmetry, while here we need just its chiral half. So, this old issue needs to be revisited. 

Finally, we would like to comment on another approach to the dual description of scattering amplitudes  
proposed in Refs.\  \cite{EdeKorSok10}. The starting point of this proposal is the correlation
function of 1/2 BPS bilinear scalar operators $\mathcal{O}(x)=\tr ( \phi^2) $. It was found that, in 
the multiple light-cone limit $x_{i,i+1}^2\to 0$, the leading asymptotic behavior of the correlation 
function $\vev{ \mathcal{O}(x_1)\ldots \mathcal{O}(x_n)}$ is given by a product of free 
propagators multiplied by a light-like bosonic Wilson loop squared.
This result was used to demonstrate the duality between the correlation function of 
bosonic operators in the light-cone limit $x_{i,i+1}^2\to 0$ and the MHV amplitudes at loop 
level. Recently, the duality between the correlation functions and amplitudes was further extended 
to non-MHV amplitudes in Refs.~\cite{EHKS}. In $\mathcal{N}=4$ SYM, the scalar operator 
$\mathcal{O}(x)$ is the lowest component of the so-called stress-tensor supermultiplet described 
by the 1/2 BPS short superfields $\mathcal{T}(x,\theta,\bar\theta)$. Then,
a natural supersymmetric generalization of the correlation function is
\begin{align}\label{TTT}
G_n = \vev{\mathcal{T}(x_1,\theta_1,\bar\theta_1)\ldots \mathcal{T}(x_n,\theta_n,\bar\theta_n)} \,.
\end{align}
This correlation function depends on both Grassmann variables, $\theta_i$ and $\bar\theta_i$, 
and it enjoys the full $\mathcal{N}=4$ superconformal symmetry to all loops.
Then, the duality relation between the correlation function and the superamplitudes \re{gen} is 
established by setting all $\bar\theta_i=0$ and  in the limit where the points $(x_i,\theta_i)$ 
coincide with the vertices of the light-like polygon $\mathcal{C}_n$  
\begin{align}\label{new}
\lim_{x_{i,i+1}^2\to 0} G_{n}(x_i,\theta_i,\bar\theta_i=0)  
\sim \lr{\sum_{k=0}^{n-4} a^{k} \widehat{\mathcal{A}}_{n;k}(\lambda_i,\tilde\lambda_i,\eta_i;a)}^2\,,
\end{align}
where the two sets of variables $(x_i,\theta_i)$ and $(\lambda_i,\tilde\lambda_i,\eta_i)$ 
are related to each other through \re{super-mom} and the proportionality factor is given 
by the product of $n$ consecutive scalar propagators $1/(x_{12}^2\ldots x_{n1}^2)$.
This relation has been checked in Ref.~\cite{EHKS} 
for $n=4,5,6$ amplitudes at tree- and one-loop level, as well as for the NMHV tree-level amplitudes general $n$. Unlike the supersymmetric Wilson loop, 
the $Q-$supersymmetry of the
correlation function in the left-hand side of \re{new} is not broken. The reason for
this is that the correlation function, viewed as a function of $x_{i,i+1}^2$, is a 
less singular object as compared with the Wilson loop. For $x_{i,i+1}^2\neq 0$ the former is well-defined in $D=4$ dimensions, while the latter suffers from
UV divergences due to the presence of (non light-like) cusps on the integration contour and, 
therefore, requires a regularization \cite{KK92}.

The situation here is very much reminiscent of the one with the operator
product expansion (OPE). Let us consider a product of two protected operators
in $\mathcal{N}=4$ SYM like $\mathcal{T}(x,\theta,\bar\theta)$. It is well-defined in $D=4$ and does not require any 
regularization as long as the operators are not null separated.
However, expanding the product of operators into the sum of local (Wilson) operators we find that the latter develop anomalous 
dimensions whose calculation requires introducing a UV regularization and whose explicit expressions depend 
(starting from two loops) on the choice of the regularization scheme. We recall 
however that each Wilson operator is accompanied by corresponding coefficient
function. It is this coefficient function that insures independence of the product
of operators on the renormalization scale as well as its scheme independence.
Coming back to the correlation function \re{TTT},  we expect that in the light-cone limit \re{new}, it 
reduces, in complete analogy with conventional OPE, to the product of the supersymmetric Wilson 
loop and a coefficient function. Along these lines, all anomalies of the supersymmetric Wilson loops
that we identified in this paper should be compensated by the coefficient function
in such a manner that their product is anomaly free.  
  
\section*{Acknowledgments}

G.K. would like to thank Luca Griguolo, Henrik Johansson and Domenico Seminara  for interesting discussions. G.K. and E.S. are grateful 
to Simon Caron-Huot, Burkhard Eden, Paul Heslop and  David Skinner for discussions. This work of A.B. was supported by the National 
Science Foundation under Grant No. PHY-0757394

\appendix

\section{Notations and conventions}
\label{NotationsConvensions}

We adopt spinor notations from Ref. \cite{BelDerKorMan03}.
We use the following conventions for raising/lowering indicies
\be
\varepsilon^{\alpha\beta} \lambda_\beta = \lambda^\alpha
\, , \qquad
\lambda^\beta \epsilon_{\beta \alpha}  = \lambda_\alpha
\, , \qquad
\widetilde\lambda_{\dot\beta} \varepsilon^{\dot\beta\dot\alpha}  = \widetilde\lambda^{\dot\alpha}
\, , \qquad
\epsilon_{\dot\alpha\dot\beta} \widetilde\lambda^{\dot\beta}  = \widetilde\lambda_{\dot\alpha}
\, , 
\ee
and notations for angle and square brackets
\be
\langle jk \rangle = \lambda^\alpha_j \lambda_{k \alpha}
\, , \qquad
[jk] = \lambda_{j \dot\alpha} \lambda_k^{\dot\alpha}
\, .
\ee
Then
\be
x^{\dot\alpha\alpha} = \sigma_\mu{}^{\dot\alpha\alpha} x^\mu
\, , \qquad
\partial^{\dot\alpha\alpha} = \sigma_\mu{}^{\dot\alpha\alpha} \partial^\mu
\, .
\ee
Such that, for instance,
\be
\partial^{\dot\alpha\alpha} x^{\dot\beta\beta} =
2 \varepsilon^{\dot\beta\dot\alpha} \varepsilon^{\alpha\beta}
\, , \qquad
\partial^{\dot\alpha\alpha} x^2 = 2 x^{\dot\alpha\alpha}
\, ,
\ee
where we used
\be
g^{\mu\nu}
\sigma_\mu{}^{\dot\alpha\alpha}
\sigma_\nu{}^{\dot\alpha\alpha}
= 2 \varepsilon^{\dot\beta\dot\alpha} \varepsilon^{\alpha\beta}
\, .
\ee
Using these definitions, one can find that
\be
x_{j,j+1} ^{\dot\alpha\alpha}  \partial_{\alpha\dot\alpha} f(x) = -2 \partial_t f (x)
\, ,
\ee
for $x = x_j - t x_{j,j+1}$.

In notations of Ref.~\cite{BelDerKorMan03}, the $\mathcal{N}=4$ SYM Lagrangian has 
the following form 
\begin{align}\notag
\mathcal{L}_{\mathcal{N}=4} =& \tr\bigg\{ -\frac12 F_{\mu\nu}F^{\mu\nu} + \frac12 D_\mu\phi^{AB} D^\mu
\bar\phi_{AB} + \frac18 g^2 [\phi^{AB},\phi^{CD}][\bar\phi_{AB},\bar\phi_{CD}]
\\
&
+i \bar{\mit\Psi}_{\dot\alpha A} \sigma_\mu^{\dot\alpha \alpha} D^\mu {\mit\Psi}_\alpha^A 
-i (D^\mu\bar{\mit\Psi}_{\dot\alpha A}) \sigma_\mu^{\dot\alpha \alpha} {\mit\Psi}_\alpha^A 
-\sqrt{2} g \, {\mit\Psi}^{\alpha A} [\bar\phi_{AB},{\mit\Psi}_\alpha^B] +\sqrt{2} g \,
\bar{\mit\Psi}_{\dot\alpha A} [\phi^{AB},\bar{\mit\Psi}^{\dot\alpha}_B]\bigg\}
\end{align}
where all fields are in the adjoint representation of the gauge group
$SU(N_c)$, and the generators are normalized as
relations
\begin{equation}
\tr (t^{a} t^{b}) = \frac{1}{2}\delta^{ab}, \qquad  t^a t^a = C_F=\frac{N_c^2-1}{2N_c}\,.
\end{equation}
It is invariant under (chiral) $Q-$supersymmetry transformations of fields
 \begin{eqnarray}
\label{N=4SUSYrules}
&&\delta_Q A^\mu
=
- i \xi^{\alpha \; A} \bar\sigma^\mu{}_{\alpha\dot\beta}
\bar{\mit\Psi}^{\dot\beta}_A
\, , \nonumber\\
&&\delta_Q \phi^{AB}
=
- i \sqrt{2}
\left\{
\xi^{\alpha \; A} {\mit\Psi}_\alpha^B - \xi^{\alpha \; B} {\mit\Psi}_\alpha^A
\right\}
\, , \nonumber\\
&&\delta_Q {\mit\Psi}^A_\alpha
=
\ft{i}2 F_{\mu\nu} \sigma^{\mu\nu}{}_\alpha{}^\beta \xi_\beta^A
+
i g [\phi^{AB} , \bar\phi_{BC}] \xi_\alpha^C
\, , \nonumber\\
&&\delta_Q \bar{\mit\Psi}_A^{\dot\alpha}
=
\sqrt{2} \left( D_\mu \bar\phi_{AB} \right)
\sigma^{\mu \, \dot\alpha\beta} \xi_\beta^B
\, .
\end{eqnarray}
However, we found it more convenient to redefine these elementary fields and transformation
parameter as follows
\be
A^{\dot\alpha\alpha} = A^\mu \sigma_\mu^{\dot\alpha\alpha}
\, , \qquad
\bar\psi^{\dot\alpha}_A = \sqrt[4]{2} \, \bar{\mit\Psi}^{\dot\alpha}_A
\, , \qquad
\psi^A_{\alpha} = \frac{1}{\sqrt[4]{2}} \, {\mit\Psi}_{\alpha}^A
\, , \qquad
F^{\alpha\beta} = \ft14 \sigma_{\mu\nu}{}^{\alpha\beta} F^{\mu\nu}
\, , \qquad
\epsilon^A_\alpha = \sqrt[4]{8} \, \xi^A_\alpha
\, .
\ee
Notice that this transformation does not change the kinetic term of the fermions. Here we defined the derivative as
$\partial^{\dot\alpha\beta} = \sigma^{\dot\alpha\beta}_\mu \partial^\mu$.

In order to simplify notations we use a uniform way of contracting the $SL(2)$ indices: undotted indices from 
upper left to lower right and dotted one lower left to upper right, that is $A^\alpha B_{\alpha\dot\alpha}$ and 
$A_{\dot\alpha} B^{\dot\alpha\alpha}$ (instead of  $A_\alpha B^{\alpha}_{\dot\alpha}$ and $A^{\dot\alpha} 
B_{\dot\alpha}^{\alpha}$), and use ket and bra notations
\begin{align}
A_\alpha \equiv \ket{A}\,,\qquad A^\alpha \equiv \bra{A} \,,\qquad
A^{\dot\alpha} \equiv |A]\,,\qquad A_{\dot\alpha} =[A|
\end{align}
In these notations, contractions of spinors take the conventional form
\begin{align}
A^\alpha B_\alpha =\vev{AB}\,,\qquad A_{\dot\alpha} B^{\dot\alpha}=[AB]
\, .
\end{align}
Then the $Q-$supersymmetric transformations take the following concise form
\ba
\label{Q-susy}
&&\delta_Q A 
=
- i \ket{\epsilon^{A}} [\bar\psi_A| 
\, , \nonumber\\
&&\delta_Q \phi^{AB}
=
- i 
\left\{\vev{\epsilon^{A} \psi ^B}-\vev{\epsilon^{B} \psi ^A} 
\right\}
\, , \nonumber\\
&&\delta_Q \ket{\psi^A}
=
i F\ket{\epsilon^A}
+
\ft{i}{2} g [\phi^{AB} , \bar\phi_{BC}]\ket{\epsilon^C}
\, , \nonumber\\
&&\delta_Q |\bar\psi_A]
=
D  \bar\phi_{AB} \ket{\epsilon^B}
\, .
\ea
The Grassmann variables have the following properties
\begin{align}\notag
& \vev{\epsilon^{A} \theta^B} = \vev{\theta^B\epsilon^{A}}  
\,,\qquad \ket{\epsilon^A} \vev{\theta^B\psi^D} + \ket{\theta^B} \vev{\psi^D\epsilon^A} 
+  \ket{\psi^D} \vev{\epsilon^A\theta^B}  =0\,,
\\[2mm]
&
\epsilon_{ABCD}\chi_1^A \chi_2^B \chi_3^C \chi_4^D = 
\epsilon_{ABCD}\chi_2^A \chi_1^B \chi_3^C \chi_4^D =\ldots=
\epsilon_{ABCD}\chi_4^A \chi_2^B \chi_3^C \chi_1^D\,.
\end{align}
Denoting 
$
\chi_1\chi_2\chi_3\chi_4\equiv \epsilon_{ABCD}\chi_1^A \chi_2^B \chi_3^C \chi_4^D 
$
we observe that in the last relation we can deal with $\chi-$variables 
as if they were commuting variables stripped from their $SU(4)$ indices
\begin{align}
 \chi_1\chi_2\chi_3\chi_4 = \chi_2\chi_1\chi_3\chi_4 =\ldots= \chi_4\chi_1\chi_2\chi_3\,.
\end{align}

\section{Conformal anomaly}
\label{conf-anom}

In this appendix, we elucidate the origin of the conformal anomaly of the supersymmetric
Wilson loop \re{K-W} and present its explicit one-loop calculation. 

To simplify the analysis
we shall examine the component of the supersymmetric Wilson loop $\mathcal{W}_{4;1}^{\rm MS}$ defined in \re{W4-ansatz} and \re{W4-c} in the special case where $\chi_2=\chi_3=\chi_4=0$. According to \re{MSsolutionWI}, the corresponding one-loop expression for  $\mathcal{W}_{4;1}^{\rm MS}$ takes the  form 
\begin{align} \label{spez}
\mathcal{W}_{4;1} 
&=-
\frac{g^2 C_F}{4\pi^2} \left( \frac{1}{x_{13}^2} + \frac{1}{x_{24}^2} \right) 
  \frac{\vev{24}^2}{\vev{12}^2\vev{41}^2}  \frac{\chi_1^4}{24}
\equiv W_1 + W_2 
\, .
\end{align}
Here, in the right-hand side, we split $\mathcal{W}_{4;1}^{\rm MS}$ into the sum of 
$W_1$ and $W_2$ corresponding to the two terms inside the parentheses.
Let us examine the action of the conformal boost $K^{\dot\alpha\alpha}$ on 
the one-loop expression \re{spez}.
We use $K=I P  I$ with the inversion acting on the spinors according to \re{Inv},
to get from \re{spez}
\begin{align}\notag
\delta_K W_1 &= W_1\bigg\{ 2\frac{\vev{2|x_2\kappa|4}}{\vev{24}}+2\frac{\vev{2|\kappa \, x_4|4}}{\vev{24}} - 4 \kappa\cdot (x_1+x_2) + 2\kappa\cdot (x_1+x_3)\bigg\}
\\\label{expect}
\delta_K W_2 &= W_2\bigg\{ 2\frac{\vev{2|x_2\kappa|4}}{\vev{24}}+2\frac{\vev{2|\kappa \, x_4|4}}{\vev{24}} - 4\kappa\cdot (x_1+x_2) + 2\kappa\cdot (x_2+x_4)\bigg\}
\end{align}
with $\kappa^{\dot\alpha\alpha}$ being the transformation parameter.
Combining together \re{expect} and \re{spez} we obtain the one-loop expression
for the conformal anomaly $\delta_K \mathcal{W}_{4;1}= \delta_K W_1+ \delta_K W_2$.

Let us reproduce the same result by making use of the conformal Ward identity for 
the supersymmetric Wilson loop. The analysis goes along the same lines as
in Sect.~\ref{SUSYWI}. Namely, we start with the path integral representation for the supersymmetric Wilson loop 
\re{path} and perform a conformal transformation
of the (super) coordinates combined with a compensating transformation of the fields inside the path integral.
The important difference from the supersymmetric
Ward identity discussed in Sect.~\ref{SUSYWI} is that the supersymmetric Wilson loop
$\mathcal{W}_{4}$ stays invariant under conformal transformations whereas
the dimensionally regularized $\mathcal{N}=4$ action changes by an amount
proportional to $(4-D)=2\varepsilon$ leading to
\begin{align}
K ^{\dot\alpha\alpha} \vev{\mathcal{W}_{4}} =  -{4i\varepsilon} \int d^{4-2\varepsilon} x \, x^{\dot\alpha\alpha} \vev{\mathcal{L}_{\mathcal{N}=4}(x){\mathcal{W}_{4}}}\,.
\end{align}
Here the expression in the right-hand side involves the insertion of the $\mathcal{N}=4$
Lagrangian into the supersymmetric Wilson loop. It proves convenient to introduce
into consideration the following quantity
\begin{align}\label{W-t}
\widetilde{\mathcal{W}}_4(k) = -i   \int d^{4-2\varepsilon} x \, \e^{ik x} \vev{\mathcal{L}_{\mathcal{N}=4}(x){\mathcal{W}_{4}}}\,.
\end{align}
Then, comparing the last two relations we find that  the conformal anomaly
is related to the behavior of $\widetilde{\mathcal{W}}_4(k)$ around $k=0$
\begin{align}\label{KW}
K ^{\dot\alpha\alpha} \vev{\mathcal{W}_{4}}  =  -2 i \varepsilon \frac{\partial \widetilde{\mathcal{W}}_4(k) }{\partial \, k_{\alpha\dot\alpha}}\bigg|_{k=0}\,.
\end{align}
Moreover, the value of the same function at the origin $\widetilde{\mathcal{W}}_4(0)$ is given by the insertion of the $\mathcal{N}=4$ action into the supersymmetric Wilson loop and is related to the derivative with respect to the coupling constant
\begin{align}\label{zero1}
\widetilde{\mathcal{W}}_4(0)=  g^2 \frac{\partial  \vev{\mathcal{W}_{4}}}{\partial g^2}=
\vev{\mathcal{W}_{4}}_{{}_{\rm 1-loop}} + O(g^4)\,.
\end{align}
As follows from \re{KW},  in order for the conformal anomaly $K ^{\dot\alpha\alpha} \vev{\mathcal{W}_{4}}$ to be different from zero the derivative in the right-hand side of \re{KW} should
develop a pole at $\varepsilon=0$.

For our purposes, it suffices to compute the one-loop correction to $\widetilde{\mathcal{W}}_4(k)$ proportional to $\chi_1^4$, which we denote by $\widetilde{\mathcal{W}}_{4;1}(k)$. To this order in the coupling, the following simplifications occur. Firstly, in the expression for the bosonic
and fermion connections, Eqs.~\re{Afield} and \re{Ffield}, entering into the definition
\re{W-prod} -- \re{Bsuperfield} of  the supersymmetric Wilson, we are allowed to retain only the terms involving the gauge field $A$ and the field-strength tensor $F$. 
Moreover, the Lagrangian $\mathcal{L}_{\mathcal{N}=4}(x)$ can be replaced in \re{W-t} by the kinetic term for the  gauge field. Its net effect is to 
modify the free gluon propagator.  Going through a lengthy calculation we obtain at small $k$
\begin{align}\notag
\widetilde{\mathcal{W}}_{4;1}(k)&= W_1\left[ 1-   \frac{i}{\varepsilon}  
{\lr{ 3 \bra{2} k|2] + \frac{ \vev{24}\bra{1} k|2]}{\vev{41}}+ O(\varepsilon)} }+ O(k^2)\right] 
\\\label{W-tilde}
& 
+ W_2\left[1 +\frac{i}{\varepsilon}   {\lr{3 \bra{4} k|4] +  \frac{ \vev{42}\bra{1} k|4]}{\vev{21}}+ O(\varepsilon)} } + O(k^2)\right]\,,
\end{align}
with $W_1$ and $W_2$ defined in \re{spez}.  Notice that the terms linear in $k$
develop a pole in $\varepsilon$ thus indicating that the correlation function 
\re{W-t} is singular for $\varepsilon\to 0$.  We can easily verify using \re{zero1} that the expression \re{W-tilde} 
is in agreement with the one-loop result \re{spez}. Then, substitution of \re{W-tilde} into \re{KW} yields
 the one-loop result for the conformal anomaly
 \begin{align}
(\kappa\cdot K)   \mathcal{W}_{4;1}  =  -{W_1} 
{\lr{ 3 \bra{2} \kappa |2] + \frac{ \vev{24}\bra{1} \kappa |2]}{\vev{41}}} }
+  {W_2}  {\lr{3 \bra{4} \kappa |4] +  \frac{ \vev{42}\bra{1} \kappa |4]}{\vev{21}}} } \,.
\end{align}
It is now straightforward to confirm that this expression coincides with the expected
result for the one-loop conformal anomaly $\delta_K \mathcal{W}_{4;1}  =\delta_K W_1 +\delta_K W_2$ 
given by \re{expect}.


\end{document}